\documentclass[aps,prd,showpacs,superscriptaddress]{revtex4} 

\usepackage{setspace}
\usepackage{graphicx}
\usepackage{natbib}
\usepackage{slashed}

\newcommand{\ptjet}{  p_{\rm  T}^{\rm  jet}}

\newcommand{\ptjetcor}{  p_{\rm  T, \rm cor}^{\rm  jet}}
\newcommand{\kt}{k_{\rm T}}
\newcommand{\yjet}{  y^{\rm  jet}}

\newcommand{\ptjetcal}{  p_{\rm  T, \rm cal}^{\rm  jet}}
\newcommand{\ptjetlead}{  p_{\rm  T,\rm cal}^{\rm  \ jet1}}
\newcommand{\yjetcal}{  y^{\rm  jet}_{\rm  cal}}
\newcommand{\fjetcal}{  \phi^{\rm  jet}_{\rm  cal}}
\newcommand{\ptjettest}{  p_{\rm  T,\rm cal}^{\rm  test \ jet}}
\newcommand{\ptjetref}{  p_{\rm  T,\rm cal}^{\rm  ref. \ jet}}

\newcommand{\ptjethad}{  p_{\rm  T,\rm had}^{\rm  jet}}
\newcommand{\yjethad}{  y^{\rm  jet}_{\rm  had}}
\newcommand{\fjethad}{  \phi^{\rm  jet}_{\rm  had}}
\newcommand{\deltapt}{  \delta_{p_{\rm  T}}^{\rm  mi}}
\newcommand{\linst}{\mathcal{L}^{\rm inst}}
\newcommand{\resoljet}{\sigma_{p_{\rm T}^{\rm jet}}}
\newcommand{\resoljetdata}{\sigma_{p_{\rm T}^{\rm jet}}^{\rm data}}
\newcommand{\resoljetmc}{\sigma_{p_{\rm T}^{\rm jet}}^{\rm mc}}
\newcommand{\ETM}       {E_T\hspace{-2.4ex}/\hspace{1.2ex}}

\begin{document}



\title{Measurement of the Inclusive Jet Cross Section using the {\boldmath $k_{\rm T}$} algorithm \\  in
{\boldmath $p\overline{p}$} Collisions at
{\boldmath $\sqrt{s}$} = 1.96 TeV with the CDF II Detector}

\vspace{0.5in}


\affiliation{Institute of Physics, Academia Sinica, Taipei, Taiwan 11529, Republic of China} 
\affiliation{Argonne National Laboratory, Argonne, Illinois 60439} 
\affiliation{Institut de Fisica d'Altes Energies, Universitat Autonoma de Barcelona, E-08193, Bellaterra (Barcelona), Spain} 
\affiliation{Baylor University, Waco, Texas  76798} 
\affiliation{Istituto Nazionale di Fisica Nucleare, University of Bologna, I-40127 Bologna, Italy} 
\affiliation{Brandeis University, Waltham, Massachusetts 02254} 
\affiliation{University of California, Davis, Davis, California  95616} 
\affiliation{University of California, Los Angeles, Los Angeles, California  90024} 
\affiliation{University of California, San Diego, La Jolla, California  92093} 
\affiliation{University of California, Santa Barbara, Santa Barbara, California 93106} 
\affiliation{Instituto de Fisica de Cantabria, CSIC-University of Cantabria, 39005 Santander, Spain} 
\affiliation{Carnegie Mellon University, Pittsburgh, PA  15213} 
\affiliation{Enrico Fermi Institute, University of Chicago, Chicago, Illinois 60637} 
\affiliation{Comenius University, 842 48 Bratislava, Slovakia; Institute of Experimental Physics, 040 01 Kosice, Slovakia} 
\affiliation{Joint Institute for Nuclear Research, RU-141980 Dubna, Russia} 
\affiliation{Duke University, Durham, North Carolina  27708} 
\affiliation{Fermi National Accelerator Laboratory, Batavia, Illinois 60510} 
\affiliation{University of Florida, Gainesville, Florida  32611} 
\affiliation{Laboratori Nazionali di Frascati, Istituto Nazionale di Fisica Nucleare, I-00044 Frascati, Italy} 
\affiliation{University of Geneva, CH-1211 Geneva 4, Switzerland} 
\affiliation{Glasgow University, Glasgow G12 8QQ, United Kingdom} 
\affiliation{Harvard University, Cambridge, Massachusetts 02138} 
\affiliation{Division of High Energy Physics, Department of Physics, University of Helsinki and Helsinki Institute of Physics, FIN-00014, Helsinki, Finland} 
\affiliation{University of Illinois, Urbana, Illinois 61801} 
\affiliation{The Johns Hopkins University, Baltimore, Maryland 21218} 
\affiliation{Institut f\"{u}r Experimentelle Kernphysik, Universit\"{a}t Karlsruhe, 76128 Karlsruhe, Germany} 
\affiliation{High Energy Accelerator Research Organization (KEK), Tsukuba, Ibaraki 305, Japan} 
\affiliation{Center for High Energy Physics: Kyungpook National University, Taegu 702-701, Korea; Seoul National University, Seoul 151-742, Korea; and SungKyunKwan University, Suwon 440-746, Korea} 
\affiliation{Ernest Orlando Lawrence Berkeley National Laboratory, Berkeley, California 94720} 
\affiliation{University of Liverpool, Liverpool L69 7ZE, United Kingdom} 
\affiliation{University College London, London WC1E 6BT, United Kingdom} 
\affiliation{Centro de Investigaciones Energeticas Medioambientales y Tecnologicas, E-28040 Madrid, Spain} 
\affiliation{Massachusetts Institute of Technology, Cambridge, Massachusetts  02139} 
\affiliation{Institute of Particle Physics: McGill University, Montr\'{e}al, Canada H3A~2T8; and University of Toronto, Toronto, Canada M5S~1A7} 
\affiliation{University of Michigan, Ann Arbor, Michigan 48109} 
\affiliation{Michigan State University, East Lansing, Michigan  48824} 
\affiliation{Institution for Theoretical and Experimental Physics, ITEP, Moscow 117259, Russia} 
\affiliation{University of New Mexico, Albuquerque, New Mexico 87131} 
\affiliation{Northwestern University, Evanston, Illinois  60208} 
\affiliation{The Ohio State University, Columbus, Ohio  43210} 
\affiliation{Okayama University, Okayama 700-8530, Japan} 
\affiliation{Osaka City University, Osaka 588, Japan} 
\affiliation{University of Oxford, Oxford OX1 3RH, United Kingdom} 
\affiliation{University of Padova, Istituto Nazionale di Fisica Nucleare, Sezione di Padova-Trento, I-35131 Padova, Italy} 
\affiliation{LPNHE, Universite Pierre et Marie Curie/IN2P3-CNRS, UMR7585, Paris, F-75252 France} 
\affiliation{University of Pennsylvania, Philadelphia, Pennsylvania 19104} 
\affiliation{Istituto Nazionale di Fisica Nucleare Pisa, Universities of Pisa, Siena and Scuola Normale Superiore, I-56127 Pisa, Italy} 
\affiliation{University of Pittsburgh, Pittsburgh, Pennsylvania 15260} 
\affiliation{Purdue University, West Lafayette, Indiana 47907} 
\affiliation{University of Rochester, Rochester, New York 14627} 
\affiliation{The Rockefeller University, New York, New York 10021} 
\affiliation{Istituto Nazionale di Fisica Nucleare, Sezione di Roma 1, University of Rome ``La Sapienza,'' I-00185 Roma, Italy} 
\affiliation{Rutgers University, Piscataway, New Jersey 08855} 
\affiliation{Texas A\&M University, College Station, Texas 77843} 
\affiliation{Istituto Nazionale di Fisica Nucleare, University of Trieste/\ Udine, Italy} 
\affiliation{University of Tsukuba, Tsukuba, Ibaraki 305, Japan} 
\affiliation{Tufts University, Medford, Massachusetts 02155} 
\affiliation{Waseda University, Tokyo 169, Japan} 
\affiliation{Wayne State University, Detroit, Michigan  48201} 
\affiliation{University of Wisconsin, Madison, Wisconsin 53706} 
\affiliation{Yale University, New Haven, Connecticut 06520} 
\author{A.~Abulencia}
\affiliation{University of Illinois, Urbana, Illinois 61801}
\author{J.~Adelman}
\affiliation{Enrico Fermi Institute, University of Chicago, Chicago, Illinois 60637}
\author{T.~Affolder}
\affiliation{University of California, Santa Barbara, Santa Barbara, California 93106}
\author{T.~Akimoto}
\affiliation{University of Tsukuba, Tsukuba, Ibaraki 305, Japan}
\author{M.G.~Albrow}
\affiliation{Fermi National Accelerator Laboratory, Batavia, Illinois 60510}
\author{D.~Ambrose}
\affiliation{Fermi National Accelerator Laboratory, Batavia, Illinois 60510}
\author{S.~Amerio}
\affiliation{University of Padova, Istituto Nazionale di Fisica Nucleare, Sezione di Padova-Trento, I-35131 Padova, Italy}
\author{D.~Amidei}
\affiliation{University of Michigan, Ann Arbor, Michigan 48109}
\author{A.~Anastassov}
\affiliation{Rutgers University, Piscataway, New Jersey 08855}
\author{K.~Anikeev}
\affiliation{Fermi National Accelerator Laboratory, Batavia, Illinois 60510}
% ...
\collaboration{CDF Collaboration\footnote{With visitors from $^a$University of Athens, 
$^b$University of Bristol, 
$^c$University Libre de Bruxelles, 
$^d$Cornell University, 
$^e$University of Cyprus, 
$^f$University of Dublin, 
$^g$University of Edinburgh, 
$^h$University of Heidelberg, 
$^i$Universidad Iberoamericana, 
$^j$University of Manchester, 
$^k$Nagasaki Institute of Applied Science, 
$^l$University de Oviedo, 
$^m$University of London, Queen Mary and Westfield College, 
$^n$Texas Tech University, 
$^o$IFIC(CSIC-Universitat de Valencia), 
}}
\noaffiliation


\date{\today}

\begin{abstract}

We report on measurements of the inclusive jet production cross section  as a function of the 
jet transverse momentum in $p \overline{p}$ collisions at $\sqrt{s} = 1.96 \ {\rm  TeV}$,  
using the $\kt$ algorithm and a data sample corresponding to $1.0 \ \rm  fb^{-1}$ collected with the  
Collider Detector at Fermilab in Run II. The  measurements are carried out in five different jet rapidity regions 
with  $| \yjet | < 2.1$ and transverse momentum in the range \mbox{$ 54 < \ptjet < 700$~GeV/$c$}.
Next-to-leading order perturbative QCD predictions are in good agreement with the measured cross sections.

\end{abstract}



\maketitle


\section{Introduction}

The measurement of the inclusive jet cross section as a function of the 
jet transverse momentum, $\ptjet$, in $p\overline{p}$ collisions at 
$\sqrt{s} = 1.96 \ \rm   TeV$  constitutes a test 
of perturbative quantum chromodynamics (pQCD)~\cite{pqcd}. 
In Run~II of the Tevatron, measurements of the jet cross section for jets with $\ptjet$ up to
about 700~GeV/$c$ \cite{ktprl,runIIjet} have extended the $\ptjet$ range by more than  150~GeV/$c$ compared to
Run~I~\cite{runIjet,d0runI,d0kt}.
%
%
In particular, the CDF collaboration recently published
results~\cite{ktprl} on inclusive jet production using the $\kt$ algorithm~\cite{ktalgo,soper} for jets with 
$\ptjet > 54$~GeV/$c$ and rapidity~\cite{coord}  in the region $0.1 < | \yjet | < 0.7$, 
which are well described by next-to-leading order (NLO) pQCD predictions~\cite{jetrad}. 
As discussed in~\cite{ktprl}, the $\kt$ algorithm has been widely used for precise QCD measurements at both
$e^+ e^-$ and $e^\pm p$ colliders, and
makes possible a well defined  comparison to the theoretical predictions~\cite{soper}.
The pQCD calculations involve matrix elements, describing the hard interaction between partons, convoluted     
with parton density functions (PDFs)~\cite{cteq,mrst} in the proton and antiproton that require input from   
experiment. The pQCD  predictions are affected by the still-limited knowledge of the gluon 
PDF,  which translates into a large uncertainty on the theoretical cross sections at 
high $\ptjet$~\cite{ktprl,runIIjet}.
Inclusive jet cross section  measurements from Run I at the Tevatron~\cite{d0runI}, performed 
in different jet rapidity regions, have been used  to partially constrain the gluon distribution in the proton.
This article  continues the studies on jet production using the $\kt$ algorithm at the Tevatron~\cite{ktprl,d0kt} and 
presents new measurements of the inclusive jet production cross section  as a function of $\ptjet$
in five different jet rapidity regions up to $| \yjet | = 2.1$,   
based on  $1.0 \rm   \ fb^{-1}$ of CDF Run II data.  
The measurements are corrected to the hadron level~\cite{hadron} and compared to NLO pQCD predictions.

\section{Experimental setup}

\begin{figure}[tbh]
\centerline{\includegraphics[width=3.5in]{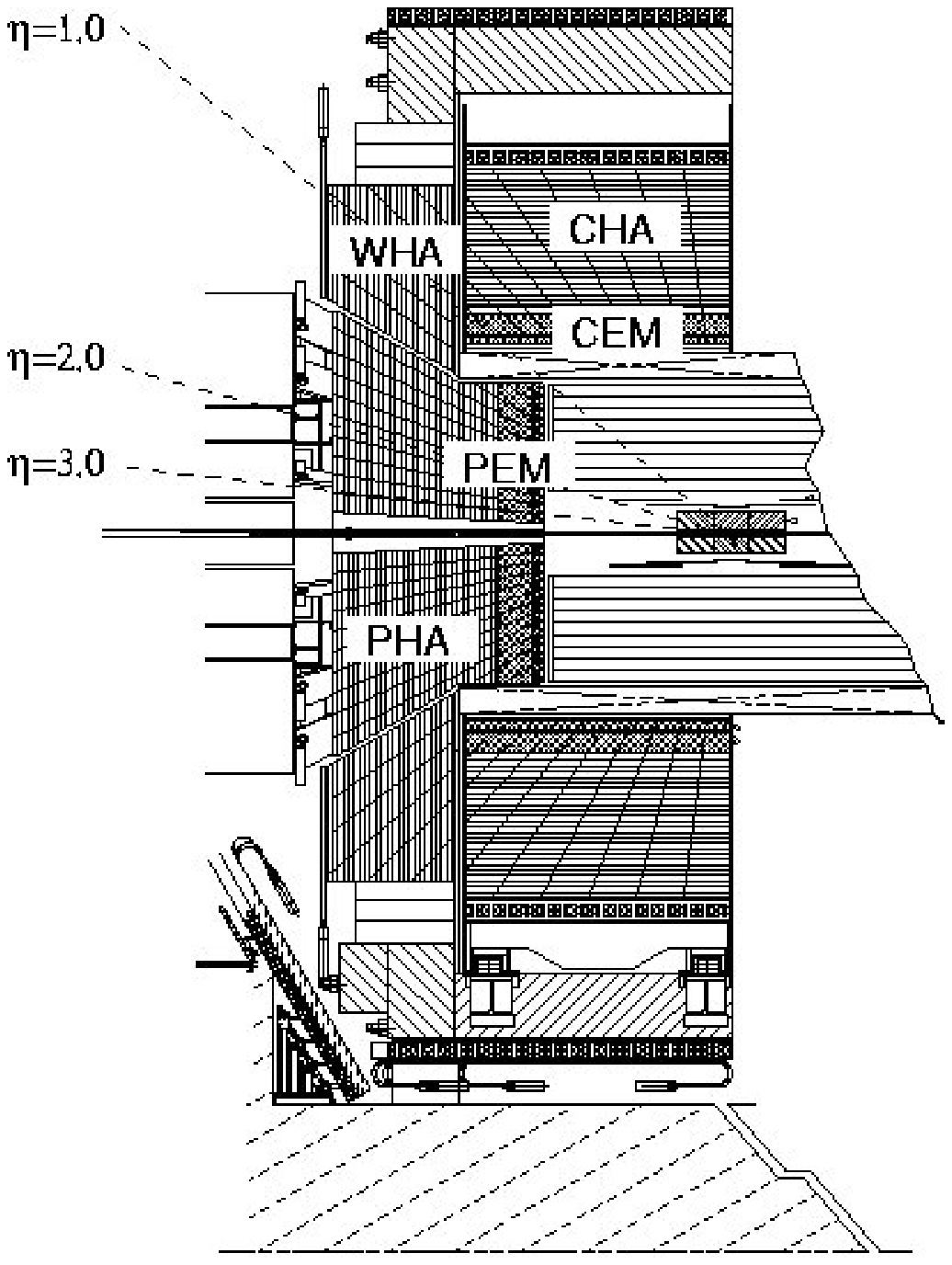}} 
\caption{Elevation view of one half of the CDF detector displaying the components of the CDF calorimeter.} 
\label{fig:cdf}
\end{figure}

The CDF II detector (see Fig.~\ref{fig:cdf}) is described in detail in \cite{cdfii}.
The sub-detectors most relevant
for this analysis are discussed briefly here. The detector has a
charged particle tracking system immersed in a 1.4~T magnetic field.
A silicon microstrip detector~\cite{silicon} provides tracking over the radial range 1.35 to 28 cm 
and covers the pseudorapidity range $|\eta| < 2$. 
A 3.1-m-long open-cell drift chamber~\cite{cot} covers the radial range from 44 to 132 cm and 
provides tracking coverage for $|\eta| < 1$. Segmented sampling calorimeters, arranged in a projective tower
geometry, surround the tracking system and measure the energy of interacting particles for
$|\eta| < 3.6$. The central barrel calorimeter~\cite{ccal} 
covers the region $|\eta| < 1$. It consists of two sections, an electromagnetic calorimeter (CEM) and a hadronic calorimeter (CHA), 
divided into 480 towers of size $0.1$ in $\eta$ and  $15^o$ in $\phi$.
The end-wall hadronic calorimeter (WHA)~\cite{wha} is behind the central barrel calorimeter in the region
$0.6 < |\eta| < 1.0$, providing forward coverage out to $|\eta| < 1.3$.
In Run II, new forward scintillator-plate calorimeters~\cite{pcal} replaced the Run I gas calorimeter system.
The new plug electromagnetic calorimeter (PEM) covers the region $1.1 < |\eta| < 3.6$,  
while the new hadronic calorimeter (PHA) provides coverage in the $1.3 < |\eta| < 3.6$ region. 
The calorimeter has gaps at $|\eta| \approx 0$ (between the two halves of the central barrel calorimeter) and at $|\eta| \approx 1.1$
(in the region between the WHA and the plug calorimeters). The measured energy resolutions for electrons in the electromagnetic 
calorimeters~\cite{ccal,pcal} are $14 \% /\sqrt{E_T} \oplus 2 \%$ (CEM) and  $16 \% /\sqrt{E} \oplus 1 \%$ (PEM), where the energies are expressed in GeV.
The single-pion energy resolutions in the hadronic calorimeters, as determined in
test-beam data~\cite{ccal,wha,pcal}, are $50 \% /\sqrt{E_T} \oplus 3 \%$ (CHA), $75 \% /\sqrt{E_T} \oplus 4\%$ (WHA) and $80 \% /\sqrt{E} \oplus 5 \%$ (PHA).
Cherenkov counters covering the $3.7 < |\eta| < 4.7$ region \cite{clc} measure the average
number of inelastic $p \overline{p}$ collisions per bunch crossing and thereby determine the beam luminosity.

\section{Jet reconstruction}

The $\kt$ algorithm~\cite{soper} is used to reconstruct jets
from the energy depositions in the 
calorimeter  towers in both data and Monte Carlo simulated events (see Section~VI).
For each calorimeter tower, the four-momenta~\cite{fn1}
of its electromagnetic and hadronic sections are summed to define 
a physics tower. First, all physics towers with transverse momentum above 0.1~GeV/$c$ are 
considered as protojets. The quantities 
\begin{equation}
  k_{\rm  T,i} = p_{\rm T,i}^2   \ \ \  {;} \ \  
  k_{\rm  T, (i,j)} =   min(p_{\rm T,i}^2,p_{\rm T,j}^2) \cdot {\Delta R_{\rm i,j}^2}/{D^2},
\end{equation}
\noindent 
are computed for each protojet and pair of protojets, respectively, where  $  p_{\rm T,i}$ denotes the transverse momentum of the 
$\rm  i^{  th}$ protojet, $\Delta R_{\rm  {i,j}}$ is the distance
($y-\phi$ space) between each pair of protojets, and $  D$ is a parameter that approximately controls the size of the jet by limiting, 
in each iteration,  
the clustering of protojets according to their spacial separation.  
All $  k_{\rm T,i}$ and $   k_{\rm  T,(i,j)}$ values are then collected into a single sorted list. 
In this list, if the smallest quantity is of the type $  k_{\rm  T,i}$, the corresponding protojet 
is promoted to be a jet
 and removed from the list. Otherwise, if the smallest quantity is of the type  $   k_{\rm  T, (i,j)}$, the protojets 
are combined into a single protojet by summing up their four-vector components. The procedure  is iterated over protojets 
until the list is empty.   
The jet transverse momentum, rapidity, and azimuthal angle are denoted as 
$\ptjetcal$, $\yjetcal$, and $\fjetcal$, respectively.

In the Monte Carlo event samples,  
the same jet algorithm is also applied to the final-state particles, considering all 
particles as protojets, to search for  jets at the hadron level.  
The resulting hadron-level jet 
variables are denoted as $\ptjethad$, $\yjethad$, and $\fjethad$. 

\section{Event selection}

Events are selected online using a three-level
trigger system~\cite{trigger} with unique
sets of selection criteria called paths. For the different trigger paths
used in this measurement, this selection is
based on the measured energy deposits in the calorimeter towers,
with different thresholds on the jet $E_T$ and different prescale factors~\cite{fn2}
(see Table~I). 
In the first-level trigger, a single trigger tower~\cite{fn3}
with $E_T$ above 5 GeV or 10 GeV, depending on the trigger path, is required.
In the second-level trigger, calorimeter clusters are
formed around the selected trigger towers. The events are required to have at least one second-level trigger cluster
with $E_T$ above a given threshold, which varies between 15 and 90 GeV for the different trigger paths. In the
third-level trigger,  jets are reconstructed using the CDF Run I cone algorithm~\cite{jetclu}, and the events
are required to have at least one jet with $E_T$ above 20 to 100 GeV.

\begin{table}[h] 
\begin{footnotesize} 
\begin{center} 
\renewcommand{\baselinestretch}{4} 
\begin{tabular}{|c|c|c|c|c|} \hline
Trigger Path & Level 1 tower $  E_{  T}$ [GeV] & Level 2 cluster  $  E_{  T}$ [GeV]& Level 3 jet $  E_{  T}$ [GeV]& eff. prescale \\ \hline\hline
JET 20 & 5  & 15 & 20 & 775 \\ \hline 
JET 50 & 5  & 40 & 50 & 34 \\ \hline 
JET 70 & 10  & 60 & 70 & 8 \\ \hline 
JET 100 & 10  & 90 & 100 & 1 \\ \hline 
\end{tabular} 
\end{center}  
\renewcommand{\baselinestretch}{1} 
\end{footnotesize}  
\label{tab:trigger}   
\caption{Summary of trigger paths, trigger thresholds and effective prescale factors employed to collect the data.}
\end{table}   

\begin{figure}[tbh]
\centerline{\includegraphics[width=4.5in]{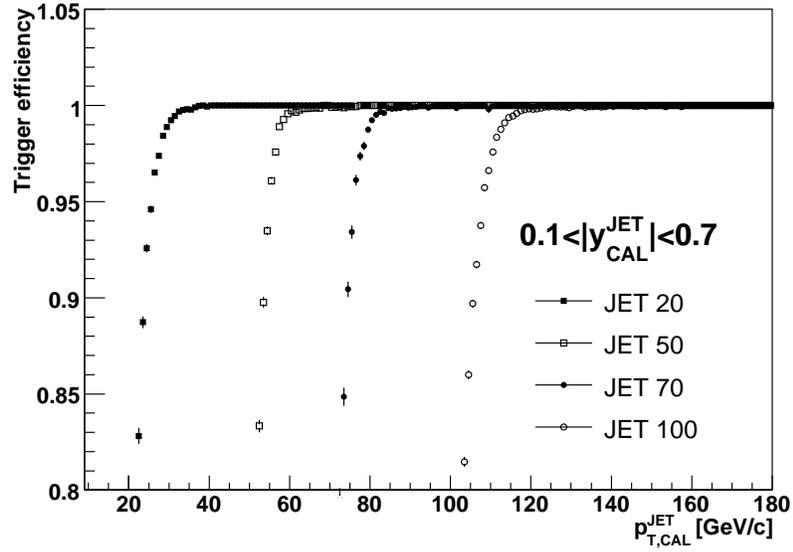}} 
\caption{Measured trigger efficiencies as a function of $\ptjetcal$  for different trigger paths and in the 
region $0.1 < |\yjetcal| < 0.7$.} 
\label{fig:trigger}
\end{figure}

\noindent
Jets are then reconstructed  using the $\kt$ algorithm, as explained in Section~III, with $  D=0.7$.
For each trigger path, the minimum $\ptjetcal$, in each $|\yjetcal|$ region,  
is chosen in such a way that the trigger selection is fully efficient. 
The efficiency for a given trigger path is obtained using events from a different trigger path with 
lower transverse energy thresholds (see Table~I). In the case of the JET 20 trigger path, 
the trigger efficiency is extracted  from 
additional control samples, which include a sample with only 
first-level trigger requirements as well as  data collected using unbiased trigger paths with no requirement on the 
energy deposits in the calorimeter towers. As an example, for 
jets in the region $0.1 < |\yjetcal| < 0.7$, Fig.~\ref{fig:trigger} shows the 
trigger efficiency as a function of $\ptjetcal$ for the different samples. 
The following selection criteria have been imposed:

\begin{enumerate} 

\item Events are required to have at least 
one reconstructed primary vertex with  $z$-position within 60~cm
of the nominal interaction point. This partially removes beam-related backgrounds and 
ensures a well-understood event-by-event jet kinematics.  

\item Events are required to have at least one jet with rapidity in the region $|\yjetcal| < 2.1$ and  
corrected $\ptjetcal$ (see Section~IX) above 54 GeV/c, which constitutes the minimum jet transverse momentum 
considered in the analysis.  The measurements are limited to jets with $|\yjetcal| < 2.1$
to avoid contributions from the $p$ and $\overline{p}$ remnants that would affect the measured  $\ptjetcal$  in the 
most forward region of the calorimeter.

\item In order to remove beam-related backgrounds and cosmic rays, the events are required to fulfill $\ETM/\sqrt{\Sigma E_T} < 
F(\ptjetlead$), where $\ETM$ denotes the missing transverse 
energy~\cite{met}  and $\Sigma E_T = \sum_i E_T^i$ is the total transverse energy of the event, as measured using 
calorimeter towers with $ E_T^i$ above 0.1~GeV. The threshold function 
$F(\ptjetlead)$ is defined as $F(\ptjet)  =  {min} (2+0.0125 \times \ptjet, 7)$, where $\ptjetlead$ is  the uncorrected 
transverse momentum of the leading jet (highest $\ptjet$) in GeV/$c$, and $F$
is in ${\rm GeV}^{1/2}$. This criterion 
preserves more than $95 \%$ of the QCD events, as determined from Monte Carlo studies (see Section~VI). 
A visual scan of events with   
$\ptjetcal > 400$~GeV/c did not show remaining backgrounds.  

\end{enumerate}

\noindent
Measurements are carried out in five different jet rapidity regions: $|\yjetcal| < 0.1$,  $0.1 < |\yjetcal| < 0.7$, 
$0.7 < |\yjetcal| < 1.1$, $1.1 < |\yjetcal| < 1.6$, and $1.6 < |\yjetcal| < 2.1$, where the 
different boundaries are chosen to reduce systematic effects coming from the 
layout of the calorimeter system.


\section{Effect of multiple {\boldmath $p\overline{p}$} interactions}

The measured $\ptjetcal$ includes contributions 
from multiple $p\overline{p}$  interactions per bunch crossing at high  
instantaneous luminosity, $\linst$. 
The data used in this measurement were collected at $\linst$  between \mbox{$0.2 \times 10^{31} {  \rm cm^{-2} s^{-1}}$} and
$16.3 \times 10^{31} {  \rm cm^{-2} s^{-1}}$ with an average of $4.1 \times 10^{31} { \rm  cm^{-2} s^{-1}}$. On  average, 
1.5 inelastic $p\overline{p}$ interactions per bunch crossing are expected. At the highest $\linst$ considered, an average of 5.9 
interactions per bunch crossing are produced.
This mainly affects the measured cross section at low 
$\ptjet$ where the contributions are sizeable. 
Multiple interactions are identified via the presence of additional primary vertices reconstructed from charged particles.
The measured jet transverse momenta are corrected for  this effect by removing a certain amount of transverse 
momentum, $\deltapt \times (N_{\rm V}-1)$, where $N_{\rm V}$ denotes the number of reconstructed primary vertices in the event 
and $\deltapt$ is determined from the data by requiring that, 
after the correction is applied, the ratio of cross sections at low and high $\linst$  does not show 
any $\ptjet$ dependence. The study is carried out separately in each $|\yjetcal|$ region, and the 
results are consistent with a 
common value $\deltapt  = 1.86 \pm 0.23$~GeV/c across the whole rapidity range.


\section{Monte Carlo simulation}

Monte Carlo simulated event samples are used to determine the response of the detector and the
correction factors to the hadron level. The generated samples are passed through a full
CDF II detector simulation (based on  {\sc geant}3~\cite{geant}, where the {\sc gflash}~\cite{gflash}
package is used to simulate the energy deposition in the calorimeters) and then reconstructed and analyzed using the same 
analysis chain as used 
for the data. 

Samples of simulated inclusive jet events have been generated with
{\sc pythia} 6.203~\cite{pythia} and {\sc herwig} 6.4~\cite{herwig} Monte Carlo generators, using
CTEQ5L~\cite{cteq5l} PDFs.
The {\sc pythia} samples have been created using a specially tuned set of parameters, denoted as
{\sc pythia-tune~a}~\cite{tunea},  that includes  enhanced contributions from initial-state
gluon radiation and  secondary parton interactions between remnants. The parameters were 
determined from dedicated studies of the underlying event 
using the CDF Run I data \cite{underlying} and  has been shown to   
properly describe the measured jet shapes in Run II~\cite{shapes}. In the case of {\sc pythia}, 
fragmentation
into hadrons is carried out using the string model \cite{string} as implemented in {\sc jetset}~\cite{jetset}, while {\sc herwig} 
implements the cluster model~\cite{cluster}.

\section{Simulation of the calorimeter response to jets}

Dedicated  studies have been performed to validate the Monte Carlo simulation of the calorimeter 
response to jets for the different $|\yjetcal|$ regions. 
Previous analyses~\cite{ktprl} for jets with $0.1 < |\yjetcal| < 0.7$  indicate that the  
simulation properly reproduces both the average $\ptjet$ and  the jet momentum resolution, $\resoljet$, as  
measured in the data. The study is performed for the rest of the $|\yjetcal|$ regions  using 
jets in the range $0.1 < |\yjetcal| < 0.7$ as a reference. An exclusive dijet sample is selected, in data and simulated events, with 
the following criteria:
\begin{enumerate}
\item Events are required to have one and only one reconstructed primary vertex with
$z$-position within 60~cm of the nominal interaction point. 
\item Events are required to have exactly two jets with $\ptjetcal > 10$~GeV/$c$, 
where one of the jets must be in the region $0.1 < |\yjetcal| < 0.7$. 
\item  $\ETM/\sqrt{\Sigma E_T} < F(\ptjetlead)$, as explained in Section~IV.
\end{enumerate} 

\begin{figure}[tbh]
\centerline{\includegraphics[width=4.5in]{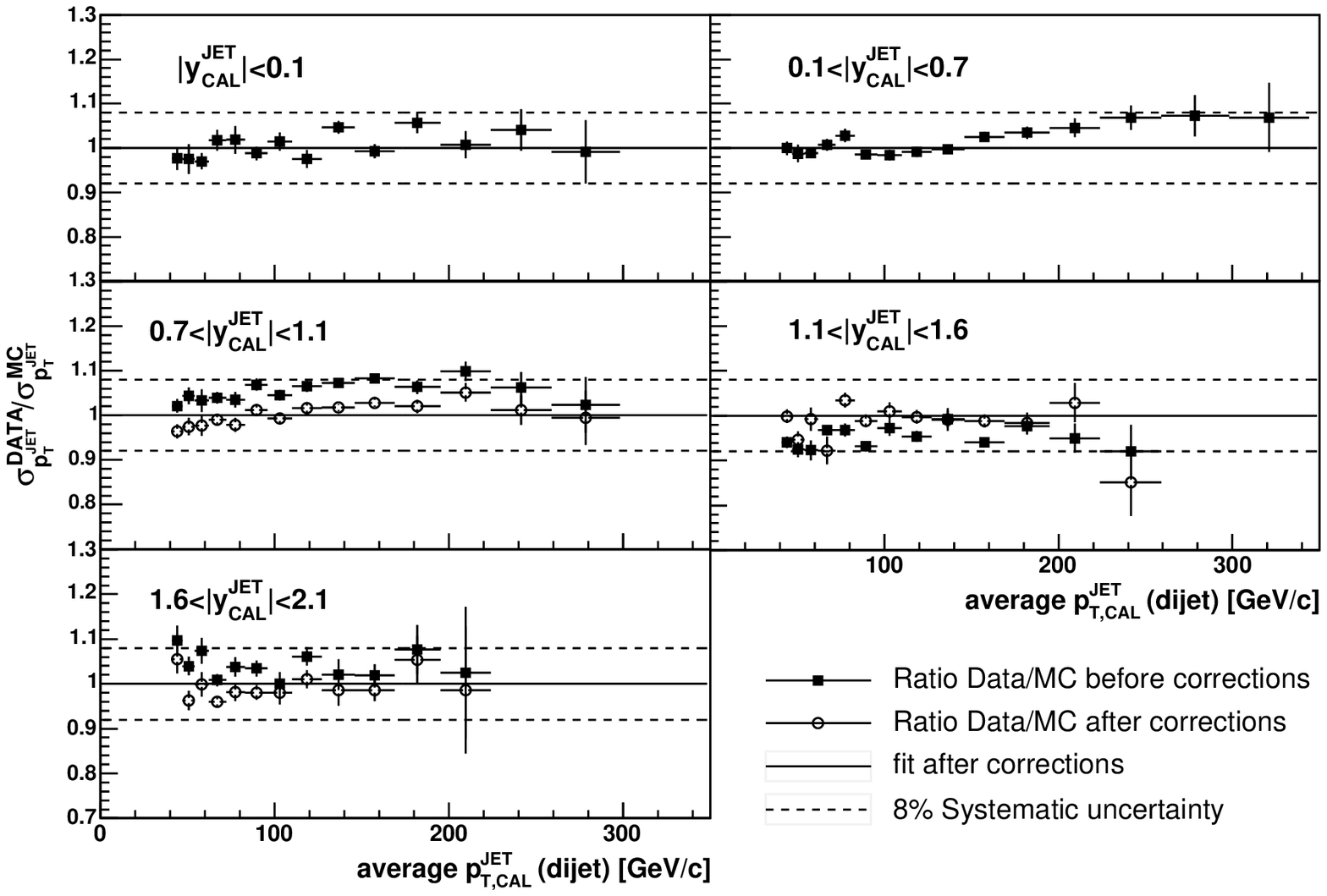}} 
\caption{Ratio $\resoljetdata/\resoljetmc$ as a function of the average $\ptjetcal$ of the dijet event, in different $|\yjetcal|$ regions,   
before (black squares) and after (open circles) corrections 
have been applied (see Section~VII). The solid lines are fits to the corrected ratios. The dashed lines indicate a $\pm 8 \%$ relative 
variation considered in the study of systematic
uncertainties.} 
\label{fig:bisector}
\end{figure}

\noindent
The bisector method~\cite{bisector} is applied to data and simulated  
exclusive dijet events to test the accuracy of the simulated 
$\resoljet$ in the detector. 
The study indicates that the  simulation systematically 
underestimates the measured $\resoljet$ by $6 \%$  and  $10 \%$ for jets in  
the regions   $0.7 < |\yjetcal| < 1.1$ and  $1.6 < |\yjetcal| < 2.1$, respectively, with 
no significant $\ptjetcal$ dependence.  
An additional smearing of the reconstructed $\ptjetcal$
is applied to the simulated events to account for this effect. 
In the region $1.1 < |\yjetcal| < 1.6$, the measured $\resoljet$ is 
overestimated by $5 \%$ in the simulation. The effect on the final result is included via 
slightly modified unfolding factors (see Section~IX). For jets  with $|\yjetcal| < 0.1$,
the  simulation  properly describes the measured $\resoljet$.
Figure~\ref{fig:bisector} shows the ratio between $\resoljet$ in data
and simulated events, $\resoljetdata/\resoljetmc$, in different
$|\yjetcal|$ regions as a function of the average $\ptjetcal$ of the dijet event. 
After corrections have been applied to the 
simulated events, data and simulation  agree. In the region $1.1 < |\yjetcal| < 1.6$, and only for 
the purpose of presentation, a $5 \%$ smearing of the reconstructed $\ptjetcal$ is applied to the data 
to show the resulting good agreement with the  uncorrected simulated resolution. 
The relative difference between data and simulated 
resolutions is conservatively 
taken to be $\pm 8 \%$ (see Fig.~\ref{fig:bisector}) over the whole range in $\ptjetcal$ and $|\yjetcal|$ in the evaluation of systematic uncertainties.

\begin{figure}[tbh]
\centerline{\includegraphics[width=4.5in]{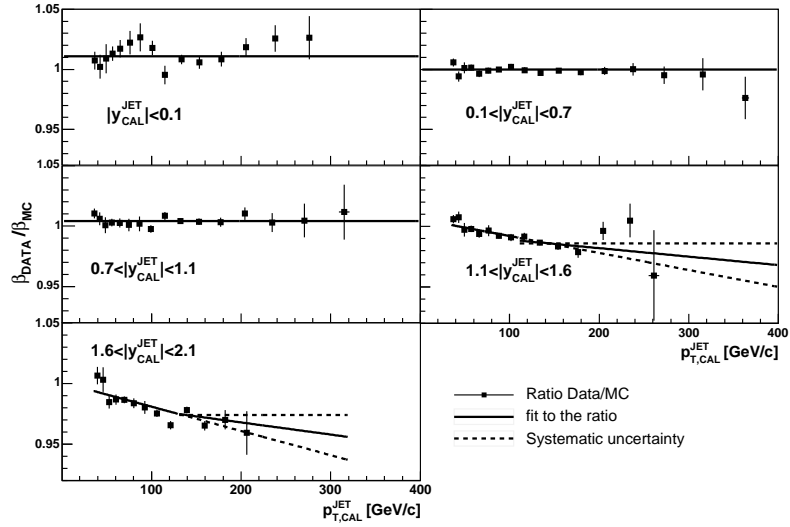}} 
\caption{
Ratio $\beta_{\rm  data}/\beta_{\rm  mc}$ as a function of $\ptjetcal$ in different 
$|\yjetcal|$ regions. The solid lines show the nominal parameterizations based on 
fits to the ratios. 
In the region $|\yjetcal| > 1.1$, the dashed lines indicate different
parameterizations used to describe the ratios at high $\ptjetcal$, and are considered in the 
study of systematic uncertainties.
} 
\label{fig:ptbalance}
\end{figure}

The average jet momentum calorimeter response in the simulation is then tested by comparing the $\ptjetcal$ balance in data and simulated exclusive 
dijet events. The variable $\beta$, defined as~\cite{fn4}
\begin{equation}
\beta = \frac{1+\langle\Delta\rangle}{1-\langle\Delta\rangle} \ , \ \  {\rm   with}  \ \ \ \Delta = \frac{\ptjettest - \ptjetref}{\ptjettest + \ptjetref}\ \ , 
\end{equation}
\noindent
is computed in data and simulated events in bins of $(\ptjettest + \ptjetref)/2 $, where $\ptjetref$ denotes the 
transverse momentum of the jet in the region  $0.1 < |\yjetcal| < 0.7$, and $\ptjettest$ is the transverse momentum of 
the jet in the $|\yjetcal|$ region under study. Figure~\ref{fig:ptbalance}
presents the ratios $\beta_{\rm  data}/\beta_{\rm  mc}$ as a function of $\ptjetcal = \ptjettest$ in the different  $|\yjetcal|$ bins. The study indicates
that small corrections are required around calorimeter gaps, $|\yjetcal| < 0.1$ and $1.1 < |\yjetcal| < 1.6$,  as well as in the
most forward region, $1.6 < |\yjetcal| < 2.1$. 
For jets with $|\yjetcal| > 1.1$, the correction shows a moderate $\ptjetcal$ dependence, and 
several parameterizations are considered to extrapolate to very high $\ptjetcal$. 
The difference observed in the final results, using different parameterizations, is included 
as part of the total systematic
uncertainty.  


\section{Reconstruction of the jet variables}

The jet reconstruction in the detector
is studied using Monte Carlo event samples, with modified jet energy response in the calorimeter,   
as described in the previous section, and pairs of jets at the calorimeter and hadron levels  
matched in $(y$ - $\phi)$ space by requiring $\sqrt{(\yjetcal - \yjethad)^2 +(\fjetcal - \fjethad)^2} < D$.
These studies
indicate that the angular variables of a jet are
reconstructed with no significant systematic shift and with a resolution better than 0.05 units 
in $y$ and $\phi$ at low $\ptjetcal$, improving as $\ptjetcal$ increases. The measured 
$\ptjetcal$ systematically underestimates that of the hadron level jet. This is attributed mainly  
to the non-compensating nature of the calorimeter~\cite{ccal}.  
For jets with $\ptjetcal$ around 50~GeV/$c$, the jet transverse
momentum is reconstructed with an average shift that varies between $-9 \%$  and $-30 \%$  and a resolution 
between $10 \%$ and $16 \%$, depending on the $|\yjetcal|$ region.
The jet reconstruction 
improves as $\ptjetcal$ increases. For jets with $\ptjetcal$ around 500~GeV/$c$, the  average shift is  
$-7 \%$ and the resolution is about $7 \%$.  


\section{Unfolding}

The measured $\ptjetcal$ distributions in the different $| \yjetcal |$ regions are  
unfolded back to the hadron level   using simulated event samples (see Section~VI), after including 
the modified jet energy response described in Section~VII. {\sc pythia-tune~a} provides a
reasonable description of the different jet and underlying event quantities, and 
is used to determine the correction factors in the unfolding procedure. In order to avoid 
any potential bias on the correction factors due to 
the particular PDF set used during the generation of the simulated samples, 
which translates into  
slightly different simulated $\ptjetcal$ distributions, the underlying $\hat{  p}_{  t}$ 
spectrum~\cite{pthat} in
{\sc pythia-tune~a}  is re-weighted until the Monte Carlo samples accurately follow
each of the measured $\ptjetcal$ distributions.~The~unfolding~is~carried out in two steps.

\begin{figure}[tbh]
\centerline{\includegraphics[width=4.5in]{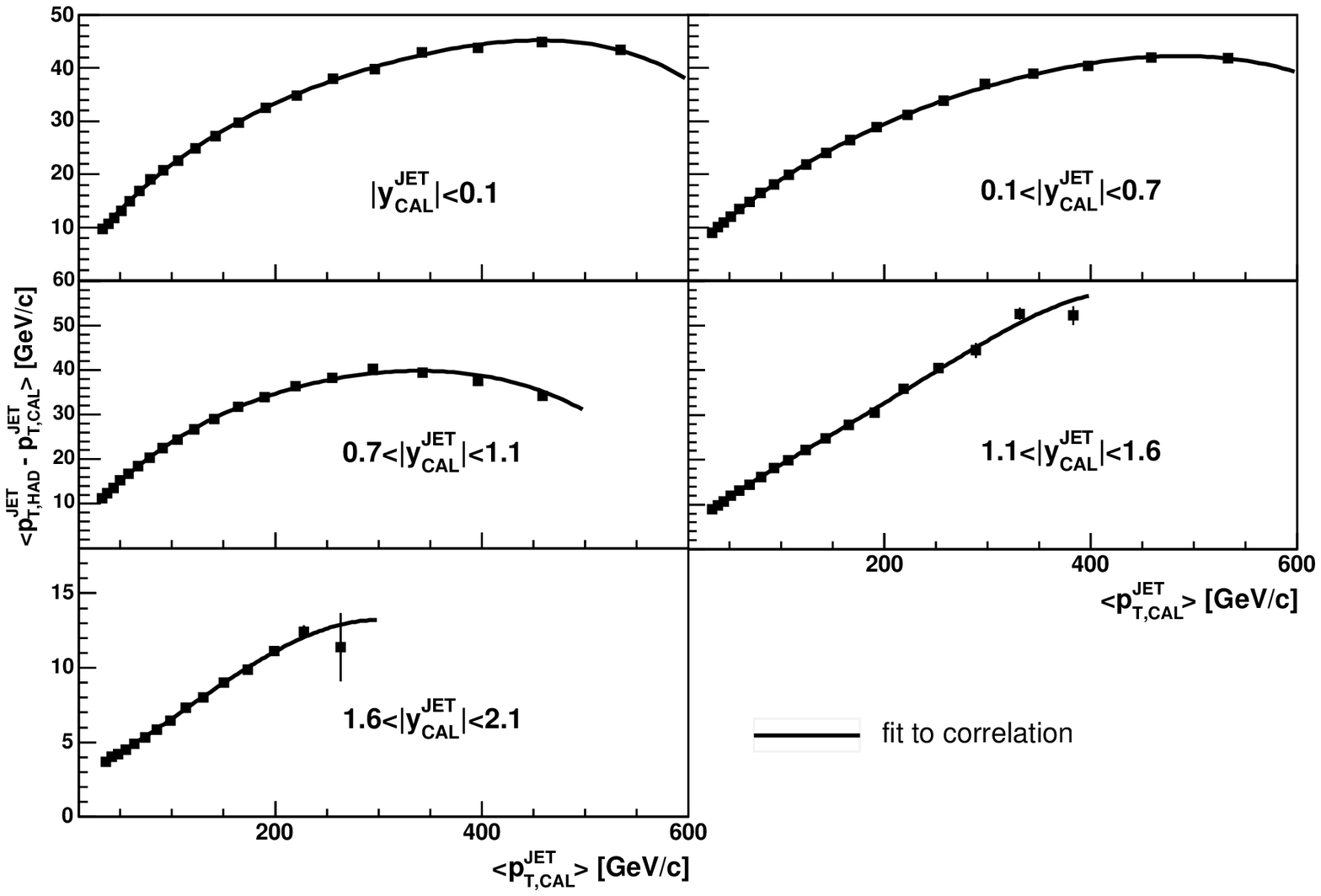}} 
\caption{Correlation  $\langle \ptjethad - \ptjetcal \rangle$ versus  $\langle\ptjetcal\rangle$, as extracted from {\sc pythia~tune~a} simulated event samples, in the different $|\yjetcal|$ regions.} 
\label{fig:ptcor}
\end{figure}

First, an average correction is computed separately in each jet rapidity region using corresponding matched pairs of jets at the calorimeter and hadron levels. 
The correlation $\langle \ptjethad - \ptjetcal \rangle$ versus  $\langle\ptjetcal\rangle$ (see  Fig.~\ref{fig:ptcor}), computed in bins of $(\ptjethad + \ptjetcal)/2$,  
is used to extract correction factors  
which are then applied to the measured jets to obtain the corrected transverse momenta, $\ptjetcor$. 
In each jet rapidity region, a cross section is  defined as
\begin{equation}
\frac{{  d}^2 \sigma}{{  d} \ptjetcor d \yjetcal} = \frac{1}{\mathcal{L}} \frac{N^{\rm  jet}_{\rm  cor}}{\Delta\ptjetcor \ \Delta\yjetcal },
\end{equation}
\noindent
where $N^{\rm  jet}_{\rm  cor}$ denotes the number of jets in a given $\ptjetcor$ bin,  
$\Delta\ptjetcor$ is the size of the bin, $ \Delta\yjetcal$ 
denotes the size of the region in $\yjetcal$, and  $\mathcal{L}$ is the integrated luminosity. $N^{\rm  jet}_{\rm  cor}$ includes event-by-event 
weights that account for trigger prescale factors, and $\Delta\ptjetcor$ is chosen according to the jet momentum resolution. 

\begin{figure}[tbh]
\centerline{\includegraphics[width=4.5in]{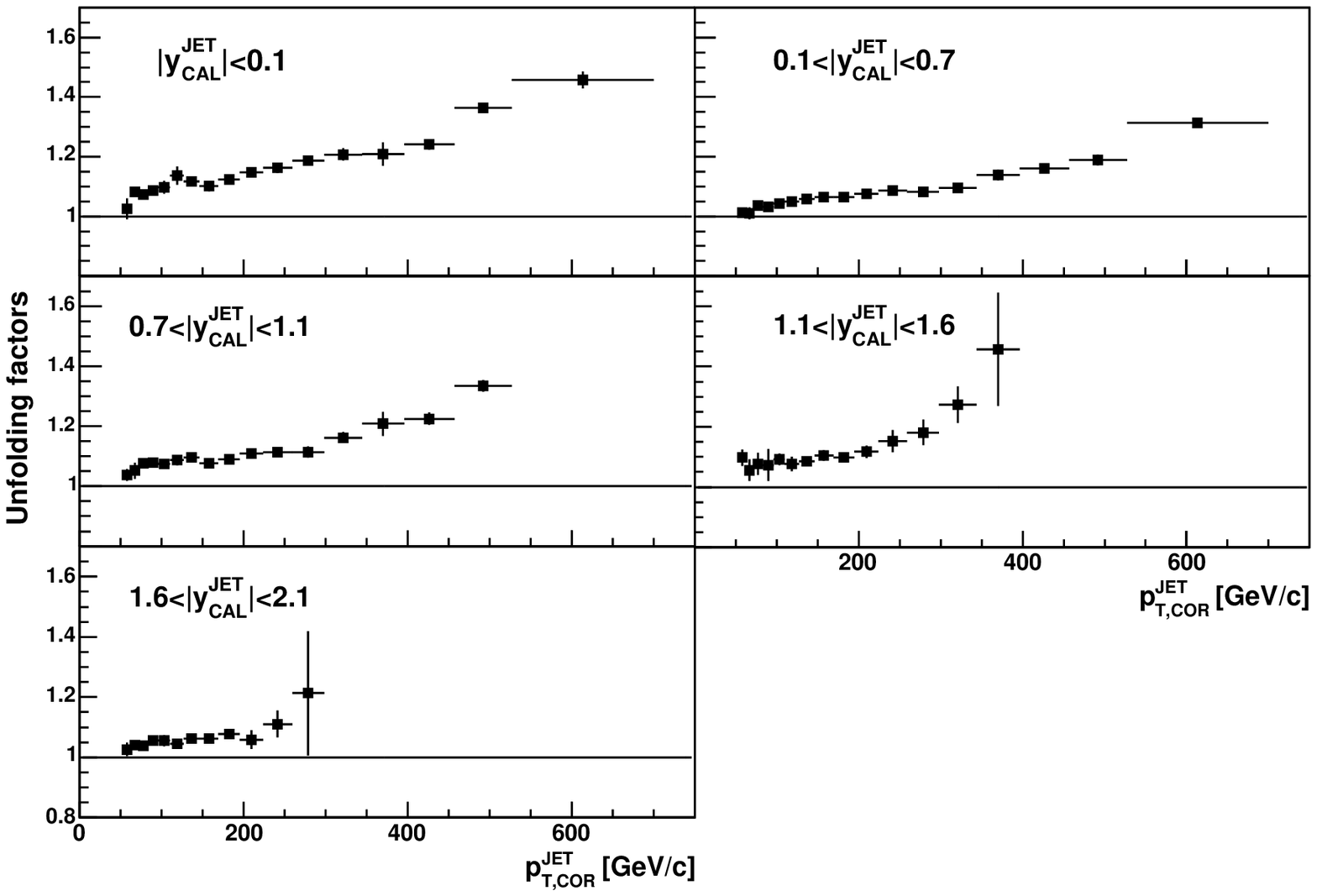}} 
\caption{Unfolding factors, $  U(\ptjetcor, \yjetcal)$, as extracted from 
{\sc pythia~tune~a} simulated event samples, as a function of $\ptjetcor$ in 
the different $|\yjetcal|$ regions.} 
\label{fig:unfolding}
\end{figure}

Second, each measurement  is corrected for acceptance
and smearing effects using a bin-by-bin unfolding procedure, which also accounts for
the efficiency of the selection criteria. The unfolding factors, defined as
\begin{equation}
  U (\ptjetcor, \yjetcal) = \frac{d^2 \sigma / d  \ptjethad d \yjethad}{d^2 \sigma / d \ptjetcor d \yjetcal},
\end{equation}
\noindent
are extracted from Monte Carlo event samples and      
applied to the measured $\ptjetcor$ distributions to obtain the final results. 
As shown in Fig.~\ref{fig:unfolding}, the factor $  U(\ptjetcor, \yjetcal)$ increases 
with $\ptjetcor$ and presents a moderate $|\yjetcal|$ dependence. At low $\ptjetcor$, the unfolding factor varies between 1.02 and 1.06 
for different rapidity regions. For jets with  $\ptjetcor$ of about 300 GeV/$c$, the factor varies between 1.1 and 1.2, and 
increases up to 1.3 - 1.4 at  very high $\ptjetcor$.  
In the region $1.1 < |\yjetcal| < 1.6$, the unfolding factor includes an additional correction, $  f_{\rm  U} (\ptjetcor)$,   
to account for the fact that the simulation overestimates the jet momentum resolution in that region 
(see Section~VII).   
The factor  $  f_{\rm  U} (\ptjetcor)$ is computed from Monte Carlo samples as the ratio between  
the $\ptjethad$ distribution
smeared using the  simulated $\resoljet$  and    
the one smeared using $\resoljet$ in data as extracted from the bisector method (see Section~VII).
The factor $  f_{\rm  U} (\ptjetcor)$ is about 1.03  and shows no 
significant $\ptjetcor$ dependence. 


\section{Systematic uncertainties}

\begin{figure}[tbh]
\centerline{\includegraphics[width=6.8in]{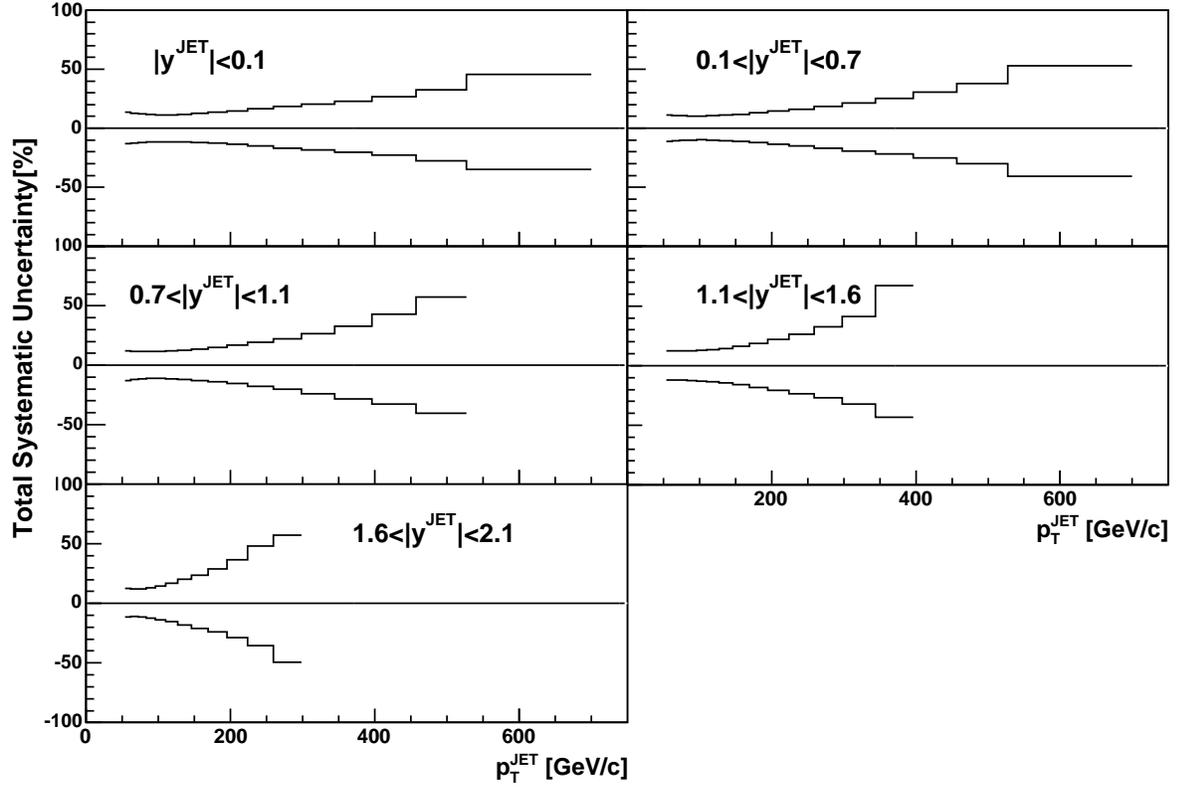}} 

\caption{
Total systematic uncertainty (in percent) on the measured inclusive differential jet cross sections as 
a function $\ptjet$  for the different $|\yjet|$ regions (see Tables~II-III). An additional 
$5.8 \%$ uncertainty on the integrated luminosity is not included. 
}
\label{fig:sys}
\end{figure}

A detailed study of the systematic uncertainties on the  measurements 
has been carried out~\cite{olga}. Tables~II-III show the different contributions to the total systematic uncertainty 
in each $\ptjet$ bin and $|\yjet|$ region:

\begin{enumerate}

\item The measured jet 
energies are varied by  $\pm 2 \%$ at low $\ptjet$ to  $\pm 2.7 \%$ at high $\ptjet$ 
to account for the uncertainty on the absolute energy scale in 
the calorimeter~\cite{jetcornim}. This introduces an uncertainty on the measured cross sections
which varies between $\pm 9 \%$ at low $\ptjet$  and ${}^{+ 61 \%}_{- 39\%}$
 at high $\ptjet$, and dominates the total systematic uncertainty on the different measurements.

\item Several sources of  systematic uncertainty on the ratio 
$\beta_{\rm  data}/\beta_{\rm  mc}$ are considered for the
different $|\yjet|$ regions: 
\begin{enumerate}
\item The uncertainty on the definition of the exclusive dijet sample in data and Monte Carlo events 
introduces a $\pm 0.5 \%$ uncertainty on the absolute energy scale 
for jets outside the region $0.1 < |\yjet| < 0.7$, which translates into an uncertainty on the cross sections between 
$\pm 2 \%$ at low $\ptjet$ and $\pm 10 \%$ at very high $\ptjet$.  
\item The use of different  $\beta_{\rm  data}/\beta_{\rm  mc}$ parameterizations for jets with 
$|\yjet|>1.1$ introduces uncertainties between  $12 \%$ and $23 \%$ at very high $\ptjet$. 
\item In the region 
 $1.1 < |\yjet| < 1.6$, 
an additional ${}^{+0\%}_{-3\%}$ uncertainty 
on the measured cross sections, independent of $\ptjet$, accounts for variations in the $\beta_{\rm  data}/\beta_{\rm  mc}$ ratio due to the  
overestimation of the jet momentum resolution in the simulated samples.    
\end{enumerate}

\item A $\pm 8 \%$ uncertainty on the jet momentum resolution introduces an uncertainty 
between $\pm 2 \%$ at low $\ptjet$ and ${}^{+ 12 \%}_{- 9 \%}$  at high $\ptjet$.    

\item The unfolding procedure is repeated using {\sc herwig} instead of {\sc pythia-tune~a} to account for the 
uncertainty on the modeling of the parton cascades and the jet fragmentation into hadrons. 
This translates into an uncertainty on the measured cross sections between $\pm 3 \%$ and $\pm 8\%$ at low $\ptjet$ that 
becomes negligible at very high $\ptjet$.

\item The unfolding procedure is also carried out using  unweighted 
{\sc pythia-tune~a}, to estimate the residual dependence on the $\ptjet$ spectra. 
This introduces an uncertainty of about $\pm 3 \%$ to $\pm 7\%$ at very high $\ptjet$, which 
becomes negligible at low $\ptjet$.
%
\item The quoted $\pm 0.23$~GeV/$c$ uncertainty 
on $\deltapt$ is taken into account.   
The maximal effect on the measured cross sections is about $\pm 2 \%$.
\item  Different sources of systematic uncertainty related to the selection criteria are considered. 
The threshold on the $z$-position of the primary vertex is varied by $\pm 5 \ \rm cm$ in data and 
simulated events. 
The lower edge of each $\ptjetcal$ bin is varied by $\pm 3 \%$ in data and simulated 
events. The $\ETM$ scale is varied by  $\pm 10 \%$ in the data.
The total effect on the measured cross sections is smaller than $1\%$ and considered negligible.
\end{enumerate}
Positive and negative deviations with respect to the nominal values in each $\ptjet$ bin are added separately in 
quadrature.  Figure~\ref{fig:sys} shows the total systematic uncertainty 
as a function of $\ptjet$ in the different $|\yjet|$ regions, where
an additional  $5.8 \%$ uncertainty on the total luminosity is  
not included. 


\section{QCD Predictions}
The measurements  are compared to parton-level NLO pQCD predictions, as computed using 
{\sc jetrad}~\cite{jetrad} with CTEQ6.1M PDFs~\cite{cteq} and the 
renormalization and factorization scales ($\mu_R$ and $\mu_F$) both set to $\mu_0 =   max (\ptjet)/2$. 
Different sources of uncertainty on the theoretical 
predictions are considered. 
The main contribution comes from the uncertainty on the PDFs and is computed using  
the Hessian method~\cite{hessian}. At low $\ptjet$, the uncertainty is about $\pm 5 \%$  and 
approximately independent of $\yjet$. The uncertainty increases as  $\ptjet$  and  $|\yjet|$ increase.
At very high $\ptjet$, the uncertainty varies between ${}^{+60 \%}_{-30 \%}$ and ${}^{+130 \%}_{-40 \%}$
for jets with $|\yjet| < 0.1$ and $1.6 < |\yjet| < 2.1$, respectively, and is  
dominated by the limited knowledge of the gluon~PDF.
An increase of $\mu_R$ and $\mu_F$ from  $\mu_0$ to $2 \mu_0$ changes the 
theoretical  predictions by only a few percent. Values  significantly smaller than $\mu_0$    
lead to unstable NLO results and are not considered.

\begin{figure}[tbh]
\centerline{\includegraphics[width=6.8in]{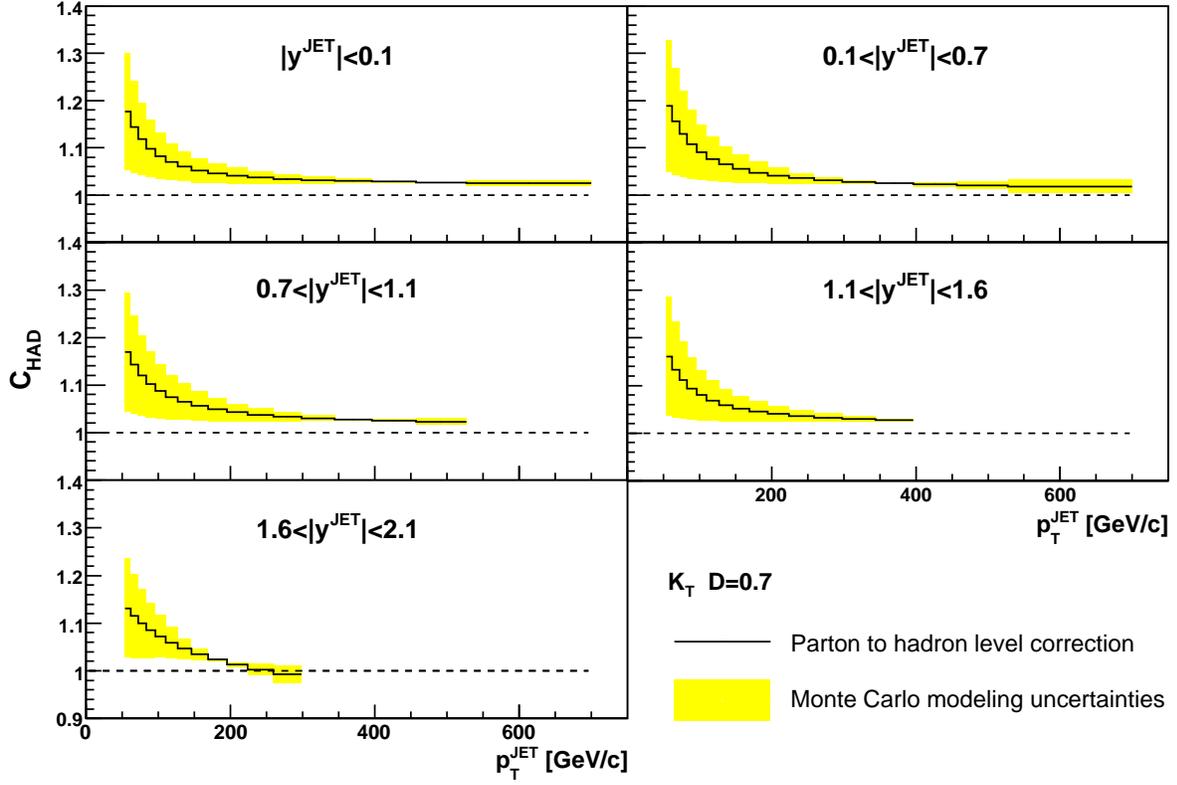}} 

\caption{Magnitude of the parton-to-hadron correction, ${ \rm C}_{  \rm HAD} (\ptjet,\yjet)$,  used to correct
the NLO pQCD predictions (see Tables~IV-V). The shaded bands indicate the quoted Monte Carlo modeling uncertainty.
}

\label{fig:chad}
\end{figure}

The theoretical predictions include a  
 correction  factor, $ {\rm C}_{  \rm HAD}(\ptjet , \yjet)$,  that approximately accounts for 
non-perturbative contributions
from  the underlying event
and fragmentation of partons into hadrons (see Fig.~\ref{fig:chad} and Tables~IV-V). 
In each jet rapidity region, $  {\rm C}_{  \rm HAD}$ is estimated, using
{\sc pythia-tune~a}, as the ratio between the nominal $\ptjethad$ distribution and the one obtained 
after removing the interactions between $p$ and $\overline{p}$ remnants and the fragmentation into hadrons 
in the Monte Carlo samples.   
The correction decreases as $\ptjet$ increases and shows a moderate $|\yjet|$ dependence. At low  
$\ptjet$, $  {\rm C}_{  \rm HAD}$
 varies between 1.18 and 1.13 as  $|\yjet|$ increases, and  it becomes of the order of 1.02 at very high $\ptjet$. 
The uncertainty on $  {\rm C}_{  \rm HAD}$ varies between  $\pm 9 \%$  and $\pm 12 \%$ at low $\ptjet$ 
and decreases  to about $\pm 1 \%$ at very high $\ptjet$, as determined 
from the difference between the parton-to-hadron correction factors obtained 
using {\sc herwig} instead of {\sc pythia-tune~a}.  

\section{Results}

\begin{figure}[tbh]
\centerline{\includegraphics[width=6.0in]{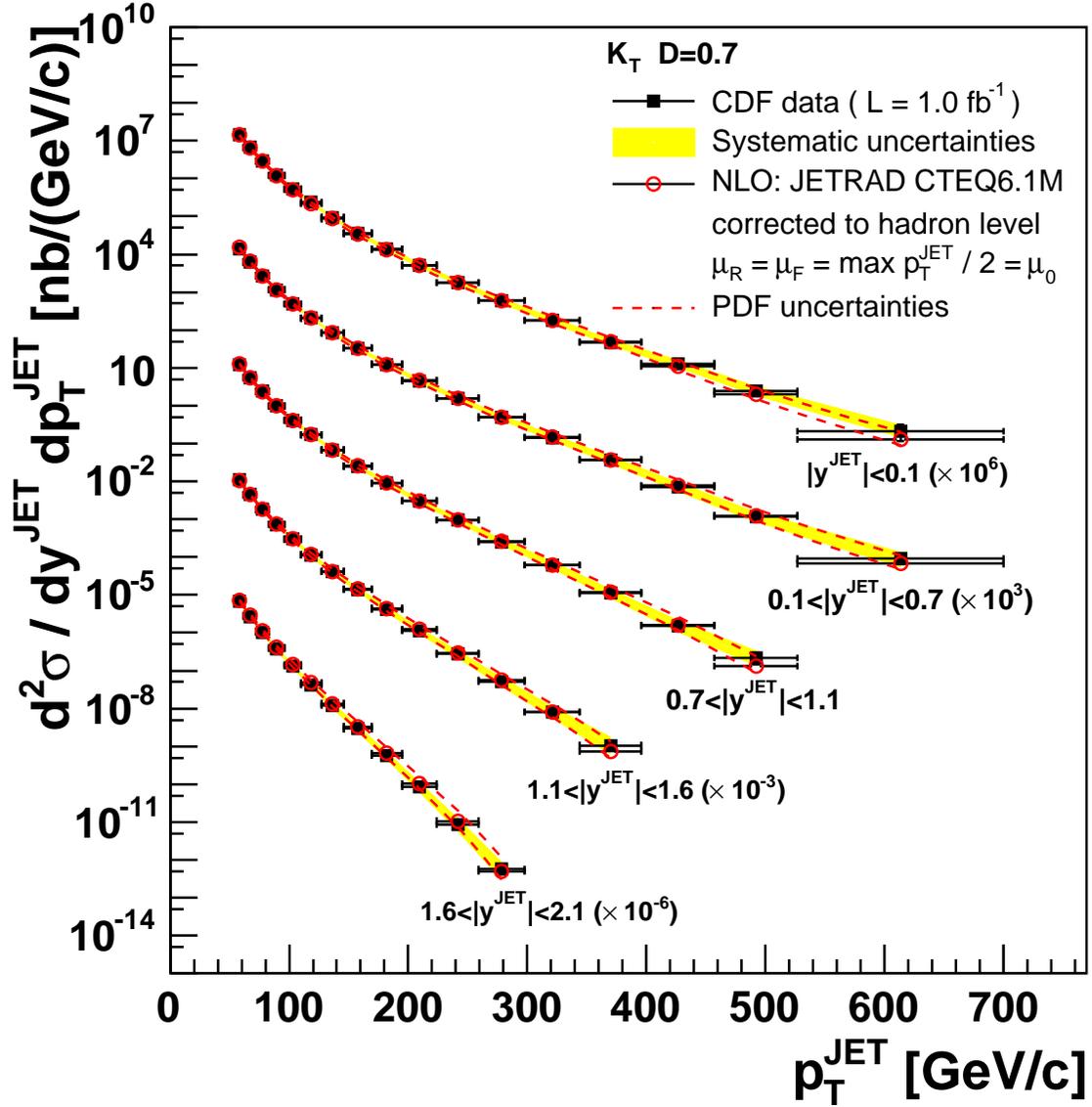}} 

\caption{Measured inclusive differential jet cross sections (black squares) as a function of $\ptjet$
for jets with $\ptjet > 54$~GeV/c 
in different $|\yjet|$ regions compared to NLO pQCD predictions (open circles).
The shaded bands show the systematic uncertainty on the measurements (see Tables~IV-V).
A $5.8 \%$ uncertainty on the integrated luminosity is
not included.
The dashed lines
indicate the PDF uncertainty on the theoretical predictions.
For presentation, the measurements in different $|\yjet|$ regions are scaled by different global factors. Factors ($\times 10^6$), 
($\times 10^3$),  ($\times 10^{-3}$), and  ($\times 10^{-6}$) are used in the regions $|\yjet| < 0.1$, $0.1 < |\yjet| < 0.7$, 
$1.1 < |\yjet| < 1.6$, and  $1.6 < |\yjet| < 2.1$, respectively.
}
\label{fig:pt}
\end{figure}

\begin{figure}[tbh]
\centerline{\includegraphics[width=7.0in]{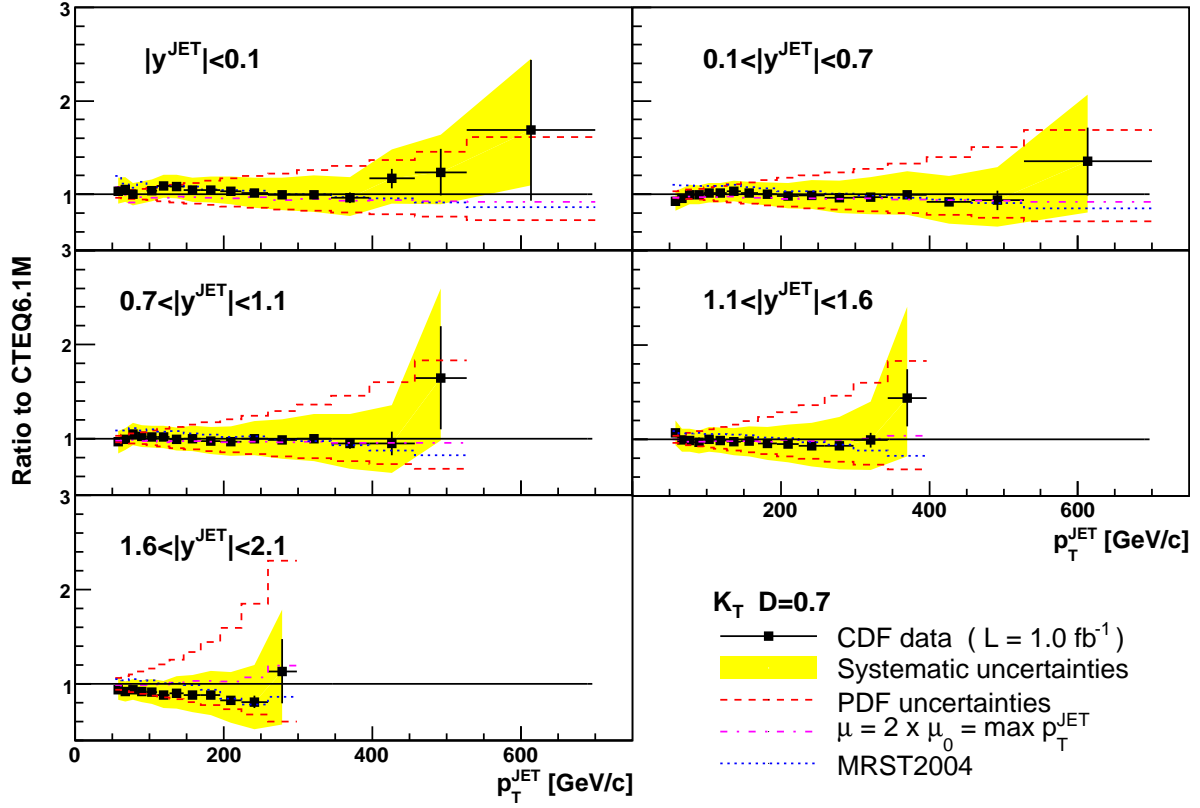}} 

\caption{
Ratio data/theory as a function of $\ptjet$ in different $|\yjet|$ regions.
The error bars (shaded bands)
show the total statistical (systematic) uncertainty on the data.
A $5.8 \%$ uncertainty on the integrated luminosity is
not included.
The dashed lines
indicate the PDF uncertainty on the theoretical predictions.
The dotted lines present the ratio of NLO pQCD predictions using MRST2004  and CTEQ6.1M PDFs.
The dotted-dashed lines show the ratios of pQCD predictions
with $2 \mu_0$ and $ \mu_0$.
} 
\label{fig:ratio}
\end{figure}

 The measured inclusive jet cross sections, 
$ \ {d^2} \sigma/ {d} \ptjet  {d} \yjet$, 
refer to  hadron-level jets,  reconstructed 
using the $\kt$ algorithm with $  D=0.7$, in the region   $\ptjet > 54$~GeV/c  and  $| \yjet |<2.1$. 
Figure~\ref{fig:pt} shows the measured cross sections as a function 
of $\ptjet$ in five different $| \yjet |$ regions compared to NLO pQCD predictions. 
The data are reported in Tables~IV-V. The measured cross sections decrease by more than seven 
to eight orders of magnitude as $\ptjet$ increases. 
Figure~\ref{fig:ratio} shows the ratio data/theory as a function of $\ptjet$ in the five different $|\yjet|$ regions. 
Good agreement is observed in the whole range in $\ptjet$ and  $\yjet$ 
between the measured cross sections  and the theoretical predictions.  In particular, 
no significant deviation from the pQCD prediction is observed for central jets at high $\ptjet$. 
The corresponding $\chi^2$ tests, relative to the nominal pQCD prediction and performed separately in each $|\yjet|$ region, 
give probabilities 
that vary between $9 \%$ and $90 \%$. A global 
$\chi^2$ test, applied to all the data points in all $|\yjet|$ regions simultaneously, gives a probability of $7 \%$. 
In both cases, a detailed treatment of correlations between systematic uncertainties was considered, as discussed in Appendix~A.
%
%
In addition, Fig.~\ref{fig:ratio} shows the ratio of pQCD predictions 
using MRST2004~\cite{mrst} and CTEQ6.1M PDF sets, 
well inside the 
theoretical and experimental uncertainties.   
The uncertainty on the measured cross sections  at high $\ptjet$,  compared to that on the 
theoretical predictions, indicates that the data presented in this article will contribute to 
a better understanding of the gluon~PDF. 

\begin{figure}[tbh]
\centerline{\includegraphics[width=7.0in]{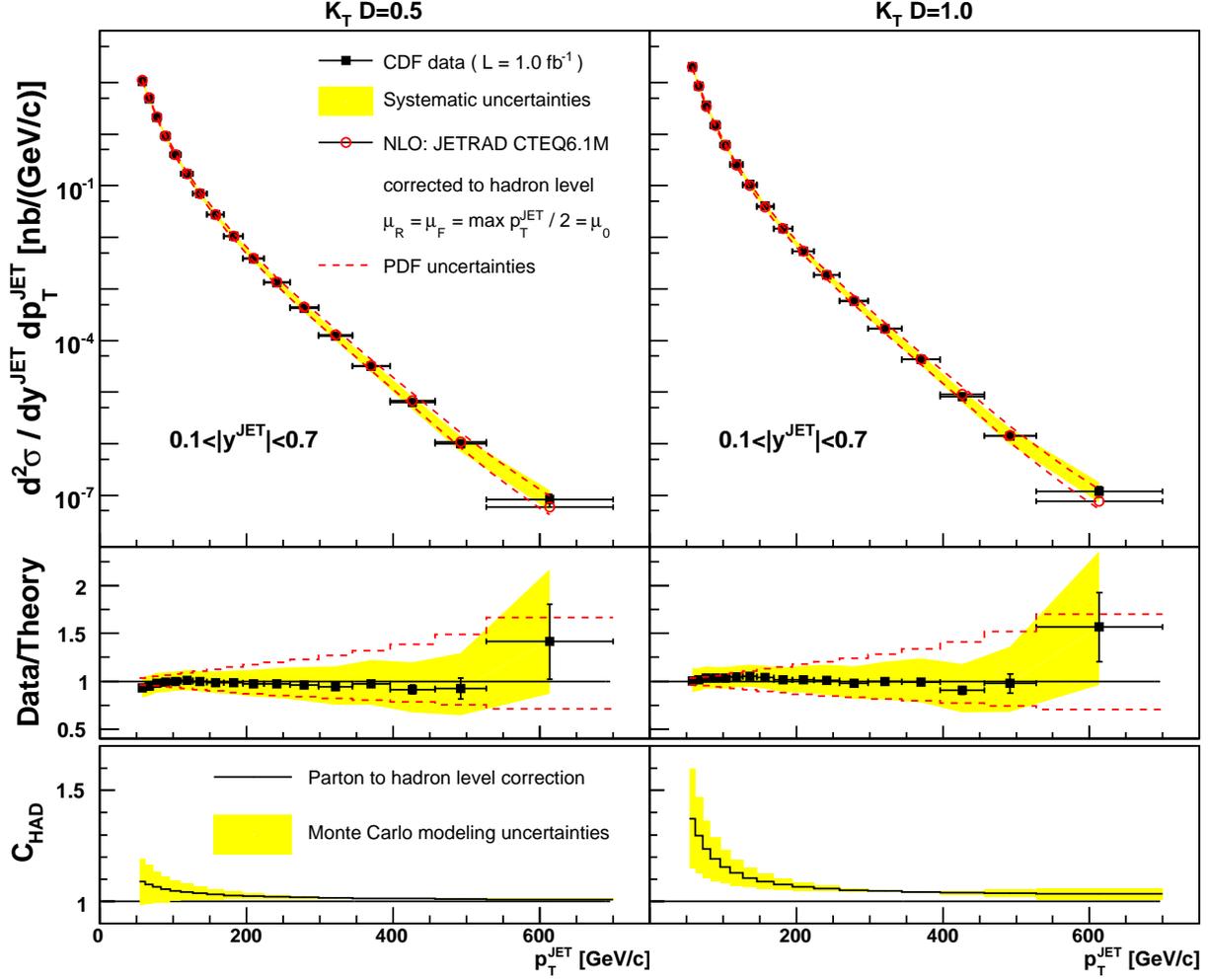}} 
\caption{(top) Measured inclusive differential jet cross sections (black squares) as a function of $\ptjet$
for jets with $\ptjet > 54$~GeV/c 
and $0.1 < |\yjet| < 0.7$ using the $\kt$ parameter $  D=0.5$ (left) and $  D=1.0$ (right), compared to NLO pQCD predictions (open circles).
The shaded bands show the total systematic uncertainty on the measurements (see Tables~VI-VII).
A $5.8 \%$ uncertainty on the integrated luminosity is
not included.
The dashed lines
indicate the PDF uncertainty on the theoretical predictions. (middle) Ratio data/theory as a function of $\ptjet$ 
for $  D=0.5$ (left) and $  D=1.0$ (right).  
(bottom) Magnitude of the parton-to-hadron corrections, $  {\rm C_{  \rm HAD}}(\ptjet)$,  used to correct
the NLO pQCD predictions for  $  D=0.5$ (left) and $  D=1.0$ (right). The shaded bands indicate the quoted Monte Carlo modeling uncertainty.} 
\label{fig:ktds}
\end{figure}

 Finally, in the region $0.1 < |\yjet| < 0.7$, the analysis is repeated using 
different values for $  D$ in 
 the $\kt$ algorithm: $  D=0.5$ and $  D=1.0$. 
In both cases, good agreement is observed between 
 the measured cross sections and the NLO pQCD predictions in the whole range in $\ptjet$ (see Fig.~\ref{fig:ktds} and  
Tables~VI-VII). The corresponding $\chi^2$ tests give probabilities of $84 \%$ and $22 \%$ for $  D=0.5$ and $  D=1.0$, respectively. 
As $  D$ decreases, the measurement is less sensitive to contributions from multiple $p\overline{p}$  
interactions per bunch crossing, and the presence and proper modeling of the underlying event. For $  D=0.5$ ($  D=1.0$), the value for $\deltapt$ becomes $1.18 \pm 0.12$ 
($3.31 \pm  0.47)$~GeV/$c$, and the parton-to-hadron correction factor applied to  the pQCD predictions 
is $  \rm C_{  \rm HAD}  = 1.1 $ ($  \rm C_{  \rm HAD} = 1.4$) at low~$\ptjet$.

\section{Summary and conclusions}

We have presented results on inclusive jet production 
in $p \overline{p}$ collisions at $\sqrt{s} = 1.96$~TeV  
for jets with transverse momentum $\ptjet > 54$~GeV/$c$ and rapidity in the region $| \yjet |<2.1$,  
using the $\kt$ algorithm and based on $1.0 \rm   \ fb^{-1}$ of CDF Run~II data. 
The measured cross sections are in agreement with NLO pQCD predictions after 
the necessary non-perturbative parton-to-hadron corrections are taken into account.
 The results reported in this article should 
contribute to a better understanding of the gluon PDF inside the proton when 
used in QCD global fits.


\section*{Acknowledgments}

We thank the Fermilab staff and the technical staffs of the participating institutions for their vital contributions. This work was supported by the U.S. Department of Energy and National Science Foundation; the Italian Istituto Nazionale di Fisica Nucleare; the Ministry of Education, Culture, Sports, Science and Technology of Japan; the Natural Sciences and Engineering Research Council of Canada; the National Science Council of the Republic of China; the Swiss National Science Foundation; the A.P. Sloan Foundation; the Bundesministerium f\"ur Bildung und Forschung, Germany; the Korean Science and Engineering Foundation and the Korean Research Foundation; the Particle Physics and Astronomy Research Council and the Royal Society, UK; the Institut National de Physique Nucleaire et Physique des Particules/CNRS; the Russian Foundation for Basic Research; the Comisi\'on Interministerial de Ciencia y Tecnolog\'{\i}a, Spain; the European Community's Human Potential Programme under contract HPRN-CT-2002-00292; and the Academy of Finland.


\clearpage



\clearpage


\clearpage


\clearpage


\clearpage


\clearpage


\clearpage


\clearpage

\clearpage



\begin{table}[htbp] 
\begin{footnotesize} 
\begin{center} 
\renewcommand{\baselinestretch}{1.2} 
\begin{tabular}{|c|c|c|c|c|c|c|c|c|} \hline
\multicolumn{9}{|c|}{\normalsize{ Systematic uncertainties [$\%$] $ (|\yjet|<0.1)$}} \\ \hline\hline
\multicolumn{1}{|c|}{$\ptjet$ [GeV/$c$]} & 
\multicolumn{1}{|c|}{jet energy scale} & 
\multicolumn{3}{|c|}{$\beta_{\rm  data}/\beta_{\rm  mc}$} & 
\multicolumn{1}{|c|}{resolution} & 
\multicolumn{1}{|c|}{unfolding} & 
\multicolumn{1}{|c|}{$\ptjet$-spectra} & 
\multicolumn{1}{|c|}{$\deltapt$} \\ 
\multicolumn{1}{|c|}{} & 
\multicolumn{1}{|c|}{} &
\multicolumn{1}{|c|}{(a)} &
\multicolumn{1}{|c|}{(b)} &
\multicolumn{1}{|c|}{(c)} &
 \multicolumn{1}{|c|}{} & 
\multicolumn{1}{|c|}{} & 
\multicolumn{1}{|c|}{} & 
\multicolumn{1}{|c|}{} \\ \hline
54 - 62  & ${}^{+ 10.3 }_{ -9.3 }$   & ${}^{+ 1.4 }_{ -2.1}$ & $-$ & $-$ & ${}^{+ 2.8 }_{ -3.0 }$   & $ \pm 8.2 $ & $\pm 1.5 $ & ${}^{+ 1.8 }_{ -1.7}$\\
62 - 72  & ${}^{+ 9.9 }_{ -9.4 }$   & ${}^{+ 1.7 }_{ -2.1}$ & $-$ & $-$ & ${}^{+ 2.8 }_{ -3.0 }$   & $ \pm 7.1 $ & $\pm 1.4 $ & ${}^{+ 1.6 }_{ -1.5}$\\
72 - 83  & ${}^{+ 9.6 }_{ -9.4 }$   & ${}^{+ 1.9 }_{ -2.1}$ & $-$ & $-$ & ${}^{+ 2.9 }_{ -3.0 }$   & $ \pm 6.2 $ & $\pm 1.3 $ & ${}^{+ 1.4 }_{ -1.3}$\\
83 - 96  & ${}^{+ 9.4 }_{ -9.5 }$   & ${}^{+ 2.1 }_{ -2.2}$ & $-$ & $-$ & ${}^{+ 2.9 }_{ -2.9 }$   & $ \pm 5.4 $ & $\pm 1.1 $ & ${}^{+ 1.3 }_{ -1.1}$\\
96 - 110  & ${}^{+ 9.5 }_{ -9.6 }$   & ${}^{+ 2.3 }_{ -2.2}$ & $-$ & $-$ & ${}^{+ 2.9 }_{ -2.9 }$   & $ \pm 4.7 $ & $\pm 1.0 $ & ${}^{+ 1.1 }_{ -1.0}$\\
110 - 127  & ${}^{+ 9.8 }_{ -9.8 }$   & ${}^{+ 2.5 }_{ -2.3}$ & $-$ & $-$ & ${}^{+ 3.0 }_{ -2.9 }$   & $ \pm 4.2 $ & $\pm 0.9 $ & ${}^{+ 1.0 }_{ -0.9}$\\
127 - 146  & ${}^{+ 10.4 }_{ -10.2 }$   & ${}^{+ 2.7 }_{ -2.4}$ & $-$ & $-$ & ${}^{+ 3.1 }_{ -2.9 }$   & $ \pm 3.7 $ & $\pm 0.8 $ & ${}^{+ 0.9 }_{ -0.8}$\\
146 - 169  & ${}^{+ 11.2 }_{ -10.8 }$   & ${}^{+ 2.8 }_{ -2.6}$ & $-$ & $-$ & ${}^{+ 3.1 }_{ -3.0 }$   & $ \pm 3.2 $ & $\pm 0.6 $ & ${}^{+ 0.8 }_{ -0.8}$\\
169 - 195  & ${}^{+ 12.4 }_{ -11.6 }$   & ${}^{+ 2.9 }_{ -2.7}$ & $-$ & $-$ & ${}^{+ 3.3 }_{ -3.0 }$   & $ \pm 2.8 $ & $\pm 0.5 $ & ${}^{+ 0.7 }_{ -0.7}$\\
195 - 224  & ${}^{+ 13.9 }_{ -12.8 }$   & ${}^{+ 3.0 }_{ -2.9}$ & $-$ & $-$ & ${}^{+ 3.4 }_{ -3.2 }$   & $ \pm 2.5 $ & $\pm 0.4 $ & ${}^{+ 0.6 }_{ -0.7}$\\
224 - 259  & ${}^{+ 15.5 }_{ -14.3 }$   & ${}^{+ 3.1 }_{ -3.1}$ & $-$ & $-$ & ${}^{+ 3.7 }_{ -3.4 }$   & $ \pm 2.2 $ & $\pm 0.3 $ & ${}^{+ 0.6 }_{ -0.6}$\\
259 - 298  & ${}^{+ 17.4 }_{ -15.9 }$   & ${}^{+ 3.3 }_{ -3.4}$ & $-$ & $-$ & ${}^{+ 4.0 }_{ -3.6 }$   & $ \pm 2.0 $ & $\pm 0.4 $ & ${}^{+ 0.5 }_{ -0.6}$\\
298 - 344  & ${}^{+ 19.5 }_{ -17.4 }$   & ${}^{+ 3.6 }_{ -3.7}$ & $-$ & $-$ & ${}^{+ 4.3 }_{ -4.0 }$   & $ \pm 1.8 $ & $\pm 0.6 $ & ${}^{+ 0.5 }_{ -0.6}$\\
344 - 396  & ${}^{+ 22.1 }_{ -19.1 }$   & ${}^{+ 4.0 }_{ -4.0}$ & $-$ & $-$ & ${}^{+ 4.8 }_{ -4.5 }$   & $ \pm 1.6 $ & $\pm 1.0 $ & ${}^{+ 0.4 }_{ -0.5}$\\
396 - 457  & ${}^{+ 25.7 }_{ -21.6 }$   & ${}^{+ 4.6 }_{ -4.4}$ & $-$ & $-$ & ${}^{+ 5.4 }_{ -5.1 }$   & $ \pm 1.4 $ & $\pm 1.8 $ & ${}^{+ 0.4 }_{ -0.5}$\\
457 - 527  & ${}^{+ 31.3 }_{ -26.3 }$   & ${}^{+ 5.3 }_{ -5.1}$ & $-$ & $-$ & ${}^{+ 6.1 }_{ -5.9 }$   & $ \pm 1.3 $ & $\pm 3.1 $ & ${}^{+ 0.3 }_{ -0.5}$\\
527 - 700  & ${}^{+ 43.7 }_{ -32.9 }$   & ${}^{+ 7.3 }_{ -6.7}$ & $-$ & $-$ & ${}^{+ 7.4 }_{ -7.3 }$   & $ \pm 1.1 $ & $\pm 7.1 $ & ${}^{+ 0.3 }_{ -0.5}$\\ \hline\hline

\multicolumn{9}{|c|}{\normalsize{ Systematic uncertainties [$\%$] $ (0.1 < |\yjet|<0.7)$}} \\ \hline\hline
\multicolumn{1}{|c|}{$\ptjet$ [GeV/$c$]} & 
\multicolumn{1}{|c|}{jet energy scale} & 
\multicolumn{3}{|c|}{$\beta_{\rm  data}/\beta_{\rm  mc}$} & 
\multicolumn{1}{|c|}{resolution} & 
\multicolumn{1}{|c|}{unfolding } & 
\multicolumn{1}{|c|}{$\ptjet$-spectra} & 
\multicolumn{1}{|c|}{$\deltapt$} \\ 
 \multicolumn{1}{|c|}{} & 
\multicolumn{1}{|c|}{} &
\multicolumn{1}{|c|}{(a)} &
\multicolumn{1}{|c|}{(b)} &
\multicolumn{1}{|c|}{(c)} &
 \multicolumn{1}{|c|}{} & 
\multicolumn{1}{|c|}{} & 
\multicolumn{1}{|c|}{} & 
\multicolumn{1}{|c|}{} \\ \hline
54 - 62  & ${}^{+ 9.5 }_{ -9.4 }$   & $-$ & $-$ & $-$ & ${}^{+ 2.2 }_{ -2.5 }$   & $ \pm 5.3 $ & $\pm 0.6 $ & ${}^{+ 1.6 }_{ -1.6}$\\
62 - 72  & ${}^{+ 9.4 }_{ -9.1 }$   & $-$ & $-$ & $-$ & ${}^{+ 2.1 }_{ -2.4 }$   & $ \pm 4.7 $ & $\pm 0.6 $ & ${}^{+ 1.5 }_{ -1.4}$\\
72 - 83  & ${}^{+ 9.4 }_{ -8.9 }$   & $-$ & $-$ & $-$ & ${}^{+ 2.1 }_{ -2.4 }$   & $ \pm 4.1 $ & $\pm 0.5 $ & ${}^{+ 1.3 }_{ -1.3}$\\
83 - 96  & ${}^{+ 9.4 }_{ -8.9 }$   & $-$ & $-$ & $-$ & ${}^{+ 2.0 }_{ -2.3 }$   & $ \pm 3.7 $ & $\pm 0.5 $ & ${}^{+ 1.2 }_{ -1.1}$\\
96 - 110  & ${}^{+ 9.6 }_{ -9.0 }$   & $-$ & $-$ & $-$ & ${}^{+ 2.0 }_{ -2.2 }$   & $ \pm 3.3 $ & $\pm 0.5 $ & ${}^{+ 1.1 }_{ -1.0}$\\
110 - 127  & ${}^{+ 10.0 }_{ -9.3 }$   & $-$ & $-$ & $-$ & ${}^{+ 1.9 }_{ -2.1 }$   & $ \pm 3.0 $ & $\pm 0.5 $ & ${}^{+ 1.0 }_{ -0.9}$\\
127 - 146  & ${}^{+ 10.6 }_{ -9.8 }$   & $-$ & $-$ & $-$ & ${}^{+ 1.9 }_{ -2.1 }$   & $ \pm 2.7 $ & $\pm 0.5 $ & ${}^{+ 0.9 }_{ -0.8}$\\
146 - 169  & ${}^{+ 11.4 }_{ -10.6 }$   & $-$ & $-$ & $-$ & ${}^{+ 1.9 }_{ -2.0 }$   & $ \pm 2.4 $ & $\pm 0.4 $ & ${}^{+ 0.8 }_{ -0.8}$\\
169 - 195  & ${}^{+ 12.6 }_{ -11.7 }$   & $-$ & $-$ & $-$ & ${}^{+ 2.0 }_{ -2.1 }$   & $ \pm 2.2 $ & $\pm 0.4 $ & ${}^{+ 0.7 }_{ -0.7}$\\
195 - 224  & ${}^{+ 14.1 }_{ -13.1 }$   & $-$ & $-$ & $-$ & ${}^{+ 2.1 }_{ -2.1 }$   & $ \pm 2.0 $ & $\pm 0.4 $ & ${}^{+ 0.7 }_{ -0.7}$\\
224 - 259  & ${}^{+ 16.0 }_{ -14.8 }$   & $-$ & $-$ & $-$ & ${}^{+ 2.2 }_{ -2.3 }$   & $ \pm 1.8 $ & $\pm 0.3 $ & ${}^{+ 0.6 }_{ -0.6}$\\
259 - 298  & ${}^{+ 18.4 }_{ -16.7 }$   & $-$ & $-$ & $-$ & ${}^{+ 2.5 }_{ -2.5 }$   & $ \pm 1.7 $ & $\pm 0.3 $ & ${}^{+ 0.6 }_{ -0.6}$\\
298 - 344  & ${}^{+ 21.3 }_{ -18.9 }$   & $-$ & $-$ & $-$ & ${}^{+ 2.8 }_{ -2.9 }$   & $ \pm 1.6 $ & $\pm 0.3 $ & ${}^{+ 0.5 }_{ -0.6}$\\
344 - 396  & ${}^{+ 25.1 }_{ -21.4 }$   & $-$ & $-$ & $-$ & ${}^{+ 3.4 }_{ -3.5 }$   & $ \pm 1.5 $ & $\pm 0.5 $ & ${}^{+ 0.5 }_{ -0.5}$\\
396 - 457  & ${}^{+ 30.3 }_{ -24.7 }$   & $-$ & $-$ & $-$ & ${}^{+ 4.1 }_{ -4.2 }$   & $ \pm 1.4 $ & $\pm 0.8 $ & ${}^{+ 0.4 }_{ -0.5}$\\
457 - 527  & ${}^{+ 37.7 }_{ -29.3 }$   & $-$ & $-$ & $-$ & ${}^{+ 5.1 }_{ -5.2 }$   & $ \pm 1.3 $ & $\pm 1.4 $ & ${}^{+ 0.4 }_{ -0.5}$\\
527 - 700  & ${}^{+ 52.3 }_{ -39.8 }$   & $-$ & $-$ & $-$ & ${}^{+ 7.3 }_{ -7.3 }$   & $ \pm 1.2 $ & $\pm 3.6 $ & ${}^{+ 0.4 }_{ -0.5}$\\ \hline\hline
\end{tabular}
\end{center}  
\renewcommand{\baselinestretch}{1} 
\end{footnotesize}  
\label{tab:sys1}   
\caption{
Systematic uncertainties (in percent) on the 
measured inclusive jet differential cross section  as a function of $\ptjet$ for jets in the regions 
$|\yjet| < 0.1$ and $0.1 < |\yjet| < 0.7$ (see Fig.~\ref{fig:sys}). The different columns follow the discussion in Section~X. 
An additional $5.8 \%$ uncertainty on the integrated luminosity is not included. 
} 
\end{table}


\begin{table}[htbp] 
\begin{footnotesize} 
\begin{center} 
\renewcommand{\baselinestretch}{1.2} 
\begin{tabular}{|c|c|c|c|c|c|c|c|c|} \hline
\multicolumn{9}{|c|}{\normalsize{ Systematic uncertainties $ [\%] \ (0.7 < |\yjet|<1.1) $}} \\ \hline\hline
\multicolumn{1}{|c|}{$\ptjet$ [GeV/$c$]} & 
\multicolumn{1}{|c|}{jet energy scale} & 
\multicolumn{3}{|c|}{$\beta_{\rm  data}/\beta_{\rm  mc}$} & 
\multicolumn{1}{|c|}{resolution} & 
\multicolumn{1}{|c|}{unfolding } & 
\multicolumn{1}{|c|}{$\ptjet$-spectra} & 
\multicolumn{1}{|c|}{$\deltapt$} \\ 
 \multicolumn{1}{|c|}{} & 
\multicolumn{1}{|c|}{} &
\multicolumn{1}{|c|}{(a)} &
\multicolumn{1}{|c|}{(b)} &
\multicolumn{1}{|c|}{(c)} &
 \multicolumn{1}{|c|}{} & 
\multicolumn{1}{|c|}{} & 
\multicolumn{1}{|c|}{} & 
\multicolumn{1}{|c|}{} \\ \hline
54 - 62  & ${}^{+ 9.2 }_{ -9.9 }$   & ${}^{+ 2.1 }_{ -2.3 }$   & $-$ & $-$ & ${}^{+ 4.0 }_{ -3.8 }$   & $ \pm 6.3 $ & $\pm 2.0 $ & ${}^{+ 1.7 }_{ -1.6}$\\
62 - 72  & ${}^{+ 9.2 }_{ -9.3 }$   & ${}^{+ 2.2 }_{ -2.3 }$   & $-$ & $-$ & ${}^{+ 3.8 }_{ -3.7 }$   & $ \pm 5.6 $ & $\pm 1.9 $ & ${}^{+ 1.5 }_{ -1.4}$\\
72 - 83  & ${}^{+ 9.2 }_{ -9.0 }$   & ${}^{+ 2.3 }_{ -2.3 }$   & $-$ & $-$ & ${}^{+ 3.7 }_{ -3.5 }$   & $ \pm 4.9 $ & $\pm 1.8 $ & ${}^{+ 1.3 }_{ -1.3}$\\
83 - 96  & ${}^{+ 9.5 }_{ -9.0 }$   & ${}^{+ 2.3 }_{ -2.3 }$   & $-$ & $-$ & ${}^{+ 3.5 }_{ -3.4 }$   & $ \pm 4.4 $ & $\pm 1.8 $ & ${}^{+ 1.2 }_{ -1.2}$\\
96 - 110  & ${}^{+ 9.9 }_{ -9.3 }$   & ${}^{+ 2.4 }_{ -2.4 }$   & $-$ & $-$ & ${}^{+ 3.4 }_{ -3.3 }$   & $ \pm 3.9 $ & $\pm 1.7 $ & ${}^{+ 1.1 }_{ -1.1}$\\
110 - 127  & ${}^{+ 10.6 }_{ -9.8 }$   & ${}^{+ 2.5 }_{ -2.5 }$   & $-$ & $-$ & ${}^{+ 3.3 }_{ -3.2 }$   & $ \pm 3.5 $ & $\pm 1.7 $ & ${}^{+ 1.0 }_{ -1.0}$\\
127 - 146  & ${}^{+ 11.5 }_{ -10.7 }$   & ${}^{+ 2.6 }_{ -2.6 }$   & $-$ & $-$ & ${}^{+ 3.3 }_{ -3.1 }$   & $ \pm 3.2 $ & $\pm 1.7 $ & ${}^{+ 0.9 }_{ -0.9}$\\
146 - 169  & ${}^{+ 12.6 }_{ -11.7 }$   & ${}^{+ 2.8 }_{ -2.7 }$   & $-$ & $-$ & ${}^{+ 3.3 }_{ -3.2 }$   & $ \pm 2.8 $ & $\pm 1.6 $ & ${}^{+ 0.8 }_{ -0.8}$\\
169 - 195  & ${}^{+ 14.1 }_{ -13.0 }$   & ${}^{+ 3.0 }_{ -2.9 }$   & $-$ & $-$ & ${}^{+ 3.4 }_{ -3.3 }$   & $ \pm 2.6 $ & $\pm 1.6 $ & ${}^{+ 0.8 }_{ -0.8}$\\
195 - 224  & ${}^{+ 15.9 }_{ -14.6 }$   & ${}^{+ 3.3 }_{ -3.2 }$   & $-$ & $-$ & ${}^{+ 3.7 }_{ -3.5 }$   & $ \pm 2.3 $ & $\pm 1.7 $ & ${}^{+ 0.7 }_{ -0.7}$\\
224 - 259  & ${}^{+ 18.1 }_{ -16.5 }$   & ${}^{+ 3.8 }_{ -3.6 }$   & $-$ & $-$ & ${}^{+ 4.1 }_{ -3.9 }$   & $ \pm 2.1 $ & $\pm 1.8 $ & ${}^{+ 0.7 }_{ -0.7}$\\
259 - 298  & ${}^{+ 21.0 }_{ -19.2 }$   & ${}^{+ 4.4 }_{ -4.1 }$   & $-$ & $-$ & ${}^{+ 4.7 }_{ -4.5 }$   & $ \pm 2.0 $ & $\pm 2.1 $ & ${}^{+ 0.6 }_{ -0.6}$\\
298 - 344  & ${}^{+ 25.2 }_{ -22.7 }$   & ${}^{+ 5.0 }_{ -4.8 }$   & $-$ & $-$ & ${}^{+ 5.6 }_{ -5.3 }$   & $ \pm 1.8 $ & $\pm 2.4 $ & ${}^{+ 0.6 }_{ -0.6}$\\
344 - 396  & ${}^{+ 31.5 }_{ -26.9 }$   & ${}^{+ 5.9 }_{ -5.6 }$   & $-$ & $-$ & ${}^{+ 6.8 }_{ -6.4 }$   & $ \pm 1.7 $ & $\pm 3.0 $ & ${}^{+ 0.6 }_{ -0.6}$\\
396 - 457  & ${}^{+ 41.3 }_{ -31.0 }$   & ${}^{+ 7.2 }_{ -6.6 }$   & $-$ & $-$ & ${}^{+ 8.3 }_{ -7.7 }$   & $ \pm 1.6 $ & $\pm 3.8 $ & ${}^{+ 0.5 }_{ -0.5}$\\
457 - 527  & ${}^{+ 55.4 }_{ -38.3 }$   & ${}^{+ 10.4 }_{ -7.7 }$   & $-$ & $-$ & ${}^{+ 10.0 }_{ -9.1 }$   & $ \pm 1.5 $ & $\pm 5.0 $ & ${}^{+ 0.5 }_{ -0.5}$\\ \hline\hline
\multicolumn{9}{|c|}{\normalsize{ Systematic uncertainties $ [\%] \ (1.1 < |\yjet|<1.6)  $}} \\ \hline\hline
\multicolumn{1}{|c|}{$\ptjet$ [GeV/$c$]} & 
\multicolumn{1}{|c|}{jet energy scale} & 
\multicolumn{3}{|c|}{$\beta_{\rm  data}/\beta_{\rm  mc}$} & 
\multicolumn{1}{|c|}{resolution} & 
\multicolumn{1}{|c|}{unfolding } & 
\multicolumn{1}{|c|}{$\ptjet$-spectra} & 
\multicolumn{1}{|c|}{$\deltapt$} \\ 
 \multicolumn{1}{|c|}{} & 
\multicolumn{1}{|c|}{} &
\multicolumn{1}{|c|}{(a)} &
\multicolumn{1}{|c|}{(b)} &
\multicolumn{1}{|c|}{(c)} &
 \multicolumn{1}{|c|}{} & 
\multicolumn{1}{|c|}{} & 
\multicolumn{1}{|c|}{} & 
\multicolumn{1}{|c|}{} \\ \hline
54 - 62  & ${}^{+ 9.4 }_{ -8.6 }$   & ${}^{+ 2.6 }_{ -2.4 }$   & $-$ & ${}^{+ 0.0 }_{ -3.0 }$   & ${}^{+ 2.9 }_{ -3.1 }$  & $ \pm 6.7 $ & $\pm 1.3 $ & ${}^{+ 1.8 }_{ -1.8}$\\
62 - 72  & ${}^{+ 9.5 }_{ -8.9 }$   & ${}^{+ 2.5 }_{ -2.4 }$   & $-$ & ${}^{+ 0.0 }_{ -3.0 }$   & ${}^{+ 2.9 }_{ -3.0 }$  & $ \pm 6.4 $ & $\pm 1.1 $ & ${}^{+ 1.6 }_{ -1.5}$\\
72 - 83  & ${}^{+ 9.8 }_{ -9.3 }$   & ${}^{+ 2.5 }_{ -2.5 }$   & $-$ & ${}^{+ 0.0 }_{ -3.0 }$   & ${}^{+ 2.9 }_{ -2.9 }$  & $ \pm 6.1 $ & $\pm 0.9 $ & ${}^{+ 1.4 }_{ -1.3}$\\
83 - 96  & ${}^{+ 10.2 }_{ -9.8 }$   & ${}^{+ 2.5 }_{ -2.6 }$   & $-$ & ${}^{+ 0.0 }_{ -3.0 }$   & ${}^{+ 2.9 }_{ -2.8 }$  & $ \pm 5.8 $ & $\pm 0.8 $ & ${}^{+ 1.3 }_{ -1.2}$\\
96 - 110  & ${}^{+ 10.9 }_{ -10.5 }$   & ${}^{+ 2.6 }_{ -2.6 }$   & $-$ & ${}^{+ 0.0 }_{ -3.0 }$   & ${}^{+ 3.0 }_{ -2.9 }$  & $ \pm 5.6 $ & $\pm 0.6 $ & ${}^{+ 1.2 }_{ -1.1}$\\
110 - 127  & ${}^{+ 11.7 }_{ -11.4 }$   & ${}^{+ 2.7 }_{ -2.8 }$   & $-$ & ${}^{+ 0.0 }_{ -3.0 }$   & ${}^{+ 3.1 }_{ -3.0 }$  & $ \pm 5.4 $ & $\pm 0.4 $ & ${}^{+ 1.1 }_{ -1.0}$\\
127 - 146  & ${}^{+ 12.8 }_{ -12.6 }$   & ${}^{+ 2.9 }_{ -3.0 }$   & $-$ & ${}^{+ 0.0 }_{ -3.0 }$   & ${}^{+ 3.4 }_{ -3.2 }$  & $ \pm 5.2 $ & $\pm 0.3 $ & ${}^{+ 1.1 }_{ -1.0}$\\
146 - 169  & ${}^{+ 14.5 }_{ -14.2 }$   & ${}^{+ 3.3 }_{ -3.3 }$   & $-$ & ${}^{+ 0.0 }_{ -3.0 }$   & ${}^{+ 3.8 }_{ -3.6 }$  & $ \pm 5.0 $ & $\pm 0.1 $ & ${}^{+ 1.0 }_{ -0.9}$\\
169 - 195  & ${}^{+ 16.9 }_{ -16.2 }$   & ${}^{+ 3.8 }_{ -3.7 }$   & $-$ & ${}^{+ 0.0 }_{ -3.0 }$   & ${}^{+ 4.3 }_{ -4.2 }$  & $ \pm 4.8 $ & $\pm 0.1 $ & ${}^{+ 1.0 }_{ -0.9}$\\
195 - 224  & ${}^{+ 20.3 }_{ -18.6 }$   & ${}^{+ 4.4 }_{ -4.2 }$   & ${}^{+ 0.7 }_{ -0.9 }$   & ${}^{+ 0.0 }_{ -3.0 }$   & ${}^{+ 5.1 }_{ -5.0 }$  & $ \pm 4.7 $ & $\pm 0.2 $ & ${}^{+ 0.9 }_{ -0.9}$\\
224 - 259  & ${}^{+ 24.7 }_{ -21.2 }$   & ${}^{+ 5.2 }_{ -5.0 }$   & ${}^{+ 2.6 }_{ -2.4 }$   & ${}^{+ 0.0 }_{ -3.0 }$   & ${}^{+ 6.2 }_{ -6.1 }$  & $ \pm 4.6 $ & $\pm 0.4 $ & ${}^{+ 0.9 }_{ -0.9}$\\
259 - 298  & ${}^{+ 29.9 }_{ -24.1 }$   & ${}^{+ 6.2 }_{ -5.9 }$   & ${}^{+ 6.3 }_{ -4.5 }$   & ${}^{+ 0.0 }_{ -3.0 }$   & ${}^{+ 7.8 }_{ -7.3 }$  & $ \pm 4.4 $ & $\pm 0.8 $ & ${}^{+ 0.9 }_{ -0.9}$\\
298 - 344  & ${}^{+ 37.2 }_{ -28.6 }$   & ${}^{+ 7.3 }_{ -7.1 }$   & ${}^{+ 12.6 }_{ -7.5 }$   & ${}^{+ 0.0 }_{ -3.0 }$   & ${}^{+ 9.8 }_{ -8.5 }$  & $ \pm 4.3 $ & $\pm 1.6 $ & ${}^{+ 0.9 }_{ -0.9}$\\
344 - 396  & ${}^{+ 61.2 }_{ -39.2 }$   & ${}^{+ 8.7 }_{ -8.3 }$   & ${}^{+ 22.7 }_{ -11.7 }$   & ${}^{+ 0.0 }_{ -3.0 }$   & ${}^{+ 12.4 }_{ -9.4 }$  & $ \pm 4.2 $ & $\pm 2.8 $ & ${}^{+ 0.9 }_{ -0.9}$\\ \hline\hline

\multicolumn{9}{|c|}{\normalsize{ Systematic uncertainties $ [\%] \ (1.6 < |\yjet|<2.1) $}} \\ \hline\hline
\multicolumn{1}{|c|}{$\ptjet$ [GeV/$c$]} & 
\multicolumn{1}{|c|}{jet energy scale} & 
\multicolumn{3}{|c|}{$\beta_{\rm  data}/\beta_{\rm  mc}$} & 
\multicolumn{1}{|c|}{resolution} & 
\multicolumn{1}{|c|}{unfolding } & 
\multicolumn{1}{|c|}{$\ptjet$-spectra} & 
\multicolumn{1}{|c|}{$\deltapt$} \\ 
\multicolumn{1}{|c|}{} & 
\multicolumn{1}{|c|}{} &
\multicolumn{1}{|c|}{(a)} &
\multicolumn{1}{|c|}{(b)} &
\multicolumn{1}{|c|}{(c)} &
 \multicolumn{1}{|c|}{} & 
\multicolumn{1}{|c|}{} & 
\multicolumn{1}{|c|}{} & 
\multicolumn{1}{|c|}{} \\ \hline
54 - 62  & ${}^{+ 11.6 }_{ -10.3 }$   & ${}^{+ 2.3 }_{ -2.1 }$   & $-$ & $-$ & ${}^{+ 1.7 }_{ -1.6 }$   & $ \pm 3.2 $ & $\pm 1.0 $ & ${}^{+ 2.1 }_{ -2.0}$\\
62 - 72  & ${}^{+ 10.9 }_{ -10.1 }$   & ${}^{+ 2.4 }_{ -2.4 }$   & $-$ & $-$ & ${}^{+ 1.6 }_{ -1.7 }$   & $ \pm 3.3 $ & $\pm 0.8 $ & ${}^{+ 1.8 }_{ -1.8}$\\
72 - 83  & ${}^{+ 11.0 }_{ -10.3 }$   & ${}^{+ 2.6 }_{ -2.6 }$   & $-$ & $-$ & ${}^{+ 1.5 }_{ -1.7 }$   & $ \pm 3.4 $ & $\pm 0.6 $ & ${}^{+ 1.7 }_{ -1.7}$\\
83 - 96  & ${}^{+ 12.0 }_{ -11.1 }$   & ${}^{+ 2.8 }_{ -2.9 }$   & $-$ & $-$ & ${}^{+ 1.5 }_{ -1.8 }$   & $ \pm 3.5 $ & $\pm 0.4 $ & ${}^{+ 1.6 }_{ -1.6}$\\
96 - 110  & ${}^{+ 13.7 }_{ -12.5 }$   & ${}^{+ 3.2 }_{ -3.2 }$   & $-$ & $-$ & ${}^{+ 1.5 }_{ -1.8 }$   & $ \pm 3.6 $ & $\pm 0.3 $ & ${}^{+ 1.5 }_{ -1.5}$\\
110 - 127  & ${}^{+ 16.2 }_{ -14.4 }$   & ${}^{+ 3.7 }_{ -3.5 }$   & $-$ & $-$ & ${}^{+ 1.6 }_{ -1.9 }$   & $ \pm 3.7 $ & $\pm 0.2 $ & ${}^{+ 1.4 }_{ -1.4}$\\
127 - 146  & ${}^{+ 19.2 }_{ -16.9 }$   & ${}^{+ 4.3 }_{ -4.0 }$   & $-$ & $-$ & ${}^{+ 1.8 }_{ -2.0 }$   & $ \pm 3.7 $ & $\pm 0.1 $ & ${}^{+ 1.4 }_{ -1.4}$\\
146 - 169  & ${}^{+ 22.8 }_{ -19.8 }$   & ${}^{+ 5.0 }_{ -4.6 }$   & $-$ & $-$ & ${}^{+ 2.1 }_{ -2.1 }$   & $ \pm 3.8 $ & $\pm 0.2 $ & ${}^{+ 1.4 }_{ -1.3}$\\
169 - 195  & ${}^{+ 27.7 }_{ -23.0 }$   & ${}^{+ 6.0 }_{ -5.4 }$   & ${}^{+ 1.3 }_{ -0.9 }$   & $-$ & ${}^{+ 2.5 }_{ -2.3 }$   & $ \pm 3.8 $ & $\pm 0.5 $ & ${}^{+ 1.4 }_{ -1.3}$\\
195 - 224  & ${}^{+ 34.9 }_{ -26.7 }$   & ${}^{+ 7.0 }_{ -6.4 }$   & ${}^{+ 5.3 }_{ -5.6 }$   & $-$ & ${}^{+ 3.0 }_{ -2.7 }$   & $ \pm 3.8 $ & $\pm 1.1 $ & ${}^{+ 1.4 }_{ -1.3}$\\
224 - 259  & ${}^{+ 46.0 }_{ -32.4 }$   & ${}^{+ 8.1 }_{ -8.0 }$   & ${}^{+ 11.0 }_{ -11.1 }$   & $-$ & ${}^{+ 3.5 }_{ -3.3 }$   & $ \pm 3.8 $ & $\pm 2.1 $ & ${}^{+ 1.4 }_{ -1.3}$\\
259 - 298  & ${}^{+ 52.9 }_{ -44.5 }$   & ${}^{+ 9.1 }_{ -10.5 }$   & ${}^{+ 19.1 }_{ -17.5 }$   & $-$ & ${}^{+ 3.9 }_{ -4.4 }$   & $ \pm 3.8 $ & $\pm 3.7 $ & ${}^{+ 1.4 }_{ -1.3}$\\ \hline
\end{tabular}
\end{center}  
\renewcommand{\baselinestretch}{1} 
\end{footnotesize}  
\label{tab:sys2}   
\caption{
Systematic uncertainties (in percent) on the
measured inclusive jet differential cross section  as a function of $\ptjet$ for jets in the regions
$0.7 < |\yjet| < 1.1$,  $1.1 < |\yjet| < 1.6$, and $1.6 < |\yjet| < 2.1$ (see Fig.~\ref{fig:sys}).  
The different columns follow the discussion in Section~X.
An additional $5.8 \%$ uncertainty on the integrated luminosity is not included. 
}
\end{table}


\begin{table}[htbp] 
\begin{footnotesize} 
\begin{center} 
\renewcommand{\baselinestretch}{1.2} 
\begin{tabular}{|c|r|c|} \hline
\multicolumn{3}{|c|}{\normalsize{$ \frac{ {d^2} \sigma}{ {d} \ptjet  {d} \yjet} \ (|\yjet|<0.1)$}} \\ \hline\hline
$\ptjet$  &  $\sigma \pm {  (stat.)} \pm {  (sys.)}$ & $  \rm C_{  \rm HAD}$ \\  
$[{\rm GeV}/c]$ & [nb/(GeV/$c$)] & parton $\rightarrow$ hadron\\ \hline
54 - 62  & $(  14.5  \pm  0.5  {}^{+2.0 }_{ -1.9 }) \times 10^{0\ \ }$ & $ 1.177  \pm  0.124 $\\
62 - 72  & $(  6.68  \pm  0.08  {}^{+0.85 }_{ -0.84 }) \times 10^{0\ \ }$ & $ 1.144  \pm  0.097 $\\
72 - 83  & $(  2.87  \pm  0.05  {}^{+0.35 }_{ -0.34 }) \times 10^{0\ \ }$ & $ 1.119  \pm  0.077 $\\
83 - 96  & $(  1.24  \pm  0.02  {}^{+0.14 }_{ -0.14 }) \times 10^{0\ \ }$ & $ 1.098  \pm  0.061 $\\
96 - 110 & $(   5.31  \pm  0.11  {}^{+0.60 }_{ -0.61 }) \times 10^{-1}$ & $ 1.083  \pm  0.049 $\\
110 - 127  & $(  2.33  \pm  0.06  {}^{+0.27 }_{ -0.26 }) \times 10^{-1}$ & $ 1.070  \pm  0.039 $\\
127 - 146  & $(  9.36  \pm  0.12  {}^{+1.10 }_{ -1.08 }) \times 10^{-2}$ & $ 1.060  \pm  0.032 $\\
146 - 169  & $(  3.63  \pm  0.06  {}^{+0.45 }_{ -0.43 }) \times 10^{-2}$ & $ 1.052  \pm  0.026 $\\
169 - 195  & $(  1.39  \pm  0.01  {}^{+0.19 }_{ -0.18 }) \times 10^{-2}$ & $ 1.046  \pm  0.021 $\\
195 - 224  & $(  5.22  \pm  0.06  {}^{+0.77 }_{ -0.72 }) \times 10^{-3}$ & $ 1.041  \pm  0.017 $\\
224 - 259  & $(  1.79  \pm  0.03  {}^{+0.29 }_{ -0.27 }) \times 10^{-3}$ & $ 1.037  \pm  0.013 $\\
259 - 298  & $(  5.92  \pm  0.11  {}^{+1.08 }_{ -1.00 }) \times 10^{-4}$ & $ 1.034  \pm  0.010 $\\
298 - 344  & $(  1.78  \pm  0.06  {}^{+0.36 }_{ -0.33 }) \times 10^{-4}$ & $ 1.032  \pm  0.007 $\\
344 - 396  & $(  4.68  \pm  0.28  {}^{+1.08 }_{ -0.94 }) \times 10^{-5}$ & $ 1.030  \pm  0.005 $\\
396 - 457  & $(  1.29  \pm  0.12  {}^{+0.34 }_{ -0.29 }) \times 10^{-5}$ & $ 1.028  \pm  0.002 $\\
457 - 527  & $(  2.47  \pm  0.50  {}^{+0.80 }_{ -0.68 }) \times 10^{-6}$ & $ 1.027  \pm  0.001 $\\
527 - 700  & $(  2.13  \pm  0.95  {}^{+0.97 }_{ -0.75 }) \times 10^{-7}$ & $ 1.026  \pm  0.006 $\\ \hline\hline
\multicolumn{3}{|c|}{\normalsize{$ \frac{ {d^2} \sigma}{ {d} \ptjet  {d} \yjet} \ (0.1 < |\yjet|<0.7)$}} \\ \hline\hline
$\ptjet$&  $\sigma \pm {  (stat.)} \pm {  (sys.)}$ & $  \rm C_{  \rm HAD}$ \\  
$[{\rm GeV}/c]$  & [nb/(GeV/$c$)] & parton $\rightarrow$ hadron\\ \hline
54 - 62  & $(  14.0  \pm  0.20   {}^{+ 1.6 }_{ -1.6 }) \times 10^{0\ \ }$ & $ 1.188  \pm  0.140 $\\
62 - 72  & $(  6.14  \pm  0.12  {}^{+ 0.66 }_{ -0.65 }) \times 10^{0\ \ }$ & $ 1.156  \pm  0.113 $\\
72 - 83  & $(  2.69  \pm  0.02  {}^{+ 0.29 }_{ -0.27 }) \times 10^{0\ \ }$ & $ 1.129  \pm  0.091 $\\
83 - 96  & $(  1.14  \pm  0.01  {}^{+ 0.12 }_{ -0.11 }) \times 10^{0\ \ }$ & $ 1.108  \pm  0.073 $\\
96 - 110  & $(  4.90  \pm  0.04  {}^{+ 0.51 }_{ -0.48 }) \times 10^{-1}$ & $ 1.090  \pm  0.059 $\\
110 - 127  & $( 2.08  \pm  0.02  {}^{+ 0.22 }_{ -0.21 }) \times 10^{-1}$ & $ 1.076  \pm  0.047 $\\
127 - 146  & $( 8.51  \pm  0.04  {}^{+ 0.95 }_{ -0.89 }) \times 10^{-2}$ & $ 1.065  \pm  0.038 $\\
146 - 169  & $( 3.33  \pm  0.02  {}^{+ 0.40 }_{ -0.37 }) \times 10^{-2}$ & $ 1.055  \pm  0.029 $\\
169 - 195  & $( 1.23  \pm  0.01  {}^{+ 0.16 }_{ -0.15 }) \times 10^{-2}$ & $ 1.047  \pm  0.023 $\\
195 - 224  & $(  4.53  \pm  0.02  {}^{+ 0.65 }_{ -0.61 }) \times 10^{-3}$ & $ 1.041  \pm  0.017 $\\
224 - 259  & $(  1.57  \pm  0.01  {}^{+ 0.26 }_{ -0.24 }) \times 10^{-3}$ & $ 1.036  \pm  0.012 $\\
259 - 298  & $(  4.87  \pm  0.06  {}^{+ 0.91 }_{ -0.83 }) \times 10^{-4}$ & $ 1.031  \pm  0.007 $\\
298 - 344  & $(  1.43  \pm  0.02  {}^{+ 0.31 }_{ -0.27 }) \times 10^{-4}$ & $ 1.028  \pm  0.003 $\\
344 - 396  & $(  3.69  \pm  0.10  {}^{+ 0.94 }_{ -0.80 }) \times 10^{-5}$ & $ 1.025  \pm  0.001 $\\
396 - 457  & $(  7.18  \pm  0.34  {}^{+ 2.20 }_{ -1.80 }) \times 10^{-6}$ & $ 1.023  \pm  0.004 $\\
457 - 527  & $(  1.16  \pm  0.13  {}^{+ 0.44 }_{ -0.35 }) \times 10^{-6}$ & $ 1.021  \pm  0.008 $\\
527 - 700  & $(  8.97  \pm  2.40  {}^{+ 4.75 }_{ -3.64 }) \times 10^{-8}$ & $ 1.018  \pm  0.014 $\\ \hline
\end{tabular} 
\end{center}  
\renewcommand{\baselinestretch}{1} 
\end{footnotesize}  
\label{tab:table1}   
\caption{Measured inclusive jet differential cross section  as a function of $\ptjet$ for jets in the regions 
$|\yjet| < 0.1$ and $0.1 < |\yjet| < 0.7$ (see Fig.~\ref{fig:pt}). 
An additional $5.8 \%$ uncertainty on the integrated luminosity is
not included.
The parton-to-hadron
 correction
factors, $  {\rm C}_{  \rm HAD} (\ptjet,\yjet)$, are applied to the pQCD predictions (see Fig.~\ref{fig:chad}). 
}
\end{table}


\begin{table}[htbp] 
\begin{footnotesize} 
\begin{center} 
\renewcommand{\baselinestretch}{1.2} 
\begin{tabular}{|c|r|c|} \hline
\multicolumn{3}{|c|}{\normalsize{$ \frac{ {d^2} \sigma}{ {d} \ptjet  {d} \yjet} \ (0.7 < |\yjet|<1.1)$}} \\ \hline\hline
$\ptjet$&  $\sigma \pm {  (stat.)} \pm {  (sys.)}$ & $  \rm C_{  \rm HAD}$ \\  
$[{\rm GeV}/c]$  & [nb/(GeV/$c$)] & parton $\rightarrow$ hadron\\ \hline
54 - 62  & $(  12.3  \pm  0.2  {}^{+ 1.5 }_{ -1.5 }) \times 10^{0\ \ }$ & $ 1.169  \pm  0.125 $\\
62 - 72  & $(  5.48  \pm  0.14  {}^{+ 0.65 }_{ -0.65 }) \times 10^{0\ \ }$ & $ 1.143  \pm  0.103 $\\
72 - 83  & $(  2.40  \pm  0.02  {}^{+ 0.28 }_{ -0.27 }) \times 10^{0\ \ }$ & $ 1.120  \pm  0.085 $\\
83 - 96  & $(  1.00  \pm  0.01  {}^{+ 0.15 }_{ -0.11 }) \times 10^{0\ \ }$ & $ 1.102  \pm  0.070 $\\
96 - 110  & $(  4.15  \pm  0.05  {}^{+ 0.48 }_{ -0.46 }) \times 10^{-1}$ & $ 1.087  \pm  0.057 $\\
110 - 127  & $( 1.73  \pm  0.03  {}^{+ 0.21 }_{ -0.20 }) \times 10^{-1}$ & $ 1.075  \pm  0.047 $\\
127 - 146  & $(  6.83  \pm  0.05  {}^{+ 0.87 }_{ -0.82 }) \times 10^{-2}$ & $ 1.064  \pm  0.038$\\
146 - 169  & $(  2.52  \pm  0.03  {}^{+ 0.35 }_{ -0.33 }) \times 10^{-2}$ & $ 1.056  \pm  0.031 $\\
169 - 195  & $(  8.95  \pm  0.06  {}^{+ 1.36 }_{ -1.26 }) \times 10^{-3}$ & $ 1.048  \pm  0.024 $\\
195 - 224  & $(  3.04  \pm  0.02  {}^{+ 0.51 }_{ -0.47 }) \times 10^{-3}$ & $ 1.042  \pm  0.019 $\\
224 - 259  & $(  9.52  \pm  0.11  {}^{+ 1.82 }_{ -1.68 }) \times 10^{-4}$ & $ 1.037  \pm  0.014 $\\
259 - 298  & $(  2.53  \pm  0.05  {}^{+ 0.56 }_{ -0.51 }) \times 10^{-4}$ & $ 1.033  \pm  0.009 $\\
298 - 344  & $(  6.18  \pm  0.17  {}^{+ 1.64 }_{ -1.49 }) \times 10^{-5}$ & $ 1.030  \pm  0.005 $\\
344 - 396  & $(  1.11  \pm  0.07  {}^{+ 0.36 }_{ -0.31 }) \times 10^{-5}$ & $ 1.027  \pm  0.001 $\\
396 - 457  & $(  1.53  \pm  0.20  {}^{+ 0.65 }_{ -0.50 }) \times 10^{-6}$ & $ 1.025  \pm  0.003 $\\
457 - 527  & $(  2.17  \pm  0.72  {}^{+ 1.25 }_{ -0.88 }) \times 10^{-7}$ & $ 1.023  \pm  0.007 $\\ \hline\hline
\multicolumn{3}{|c|}{\normalsize{$ \frac{ {d^2} \sigma}{ {d} \ptjet  {d} \yjet} \ (1.1 < |\yjet|<1.6)$}} \\ \hline\hline
$\ptjet$&  $\sigma \pm {  (stat.)} \pm {  (sys.)}$ & $  \rm C_{  \rm HAD}$ \\  
$[{\rm GeV}/c]$  & [nb/(GeV/$c$)] & parton $\rightarrow$ hadron\\ \hline
54 - 62  & $(  11.0  \pm  0.3   {}^{+ 1.4 }_{ -1.3 }  ) \times 10^{0\ \ }$ & $ 1.160  \pm  0.125 $\\
62 - 72  & $(  4.40  \pm  0.15  {}^{+ 0.54 }_{ -0.53 }) \times 10^{0\ \ }$ & $ 1.133  \pm  0.101 $\\
72 - 83  & $(  1.82  \pm  0.06  {}^{+ 0.22 }_{ -0.22 }) \times 10^{0\ \ }$ & $ 1.111  \pm  0.081 $\\
83 - 96  & $(  7.22  \pm  0.37  {}^{+ 0.90 }_{ -0.90 }) \times 10^{-1}$ & $ 1.094  \pm  0.065 $\\
96 - 110 & $(  2.98  \pm  0.05  {}^{+ 0.38 }_{ -0.38 }) \times 10^{-1}$ & $ 1.080  \pm  0.052 $\\
110 - 127 & $( 1.14  \pm  0.03  {}^{+ 0.15 }_{ -0.15 }) \times 10^{-1}$ & $ 1.068  \pm  0.042 $\\
127 - 146  & $(  4.10  \pm  0.04  {}^{+ 0.60 }_{ -0.60 }) \times 10^{-2}$ & $ 1.059  \pm  0.034 $\\
146 - 169  & $(  1.39  \pm  0.02  {}^{+ 0.22 }_{ -0.23 }) \times 10^{-2}$ & $ 1.051  \pm  0.027 $\\
169 - 195  & $(  4.19  \pm  0.04  {}^{+ 0.78 }_{ -0.76 }) \times 10^{-3}$ & $ 1.045  \pm  0.021 $\\
195 - 224  & $(  1.15  \pm  0.02  {}^{+ 0.25 }_{ -0.24 }) \times 10^{-3}$ & $ 1.040  \pm  0.016 $\\
224 - 259  & $(  2.73  \pm  0.09  {}^{+ 0.73 }_{ -0.64 }) \times 10^{-4}$ & $ 1.036  \pm  0.012 $\\
259 - 298  & $(  5.18  \pm  0.23  {}^{+ 1.68 }_{ -1.39 }) \times 10^{-5}$ & $ 1.033  \pm  0.009 $\\
298 - 344  & $(  7.99  \pm  0.61  {}^{+ 3.31 }_{ -2.56 }) \times 10^{-6}$ & $ 1.030  \pm  0.006 $\\
344 - 396  & $(  1.05  \pm  0.22  {}^{+ 0.71 }_{ -0.45 }) \times 10^{-6}$ & $ 1.028  \pm  0.003 $\\ \hline\hline
\multicolumn{3}{|c|}{\normalsize{$ \frac{ {d^2} \sigma}{ {d} \ptjet  {d} \yjet} \ (1.6 < |\yjet|<2.1) $}} \\ \hline\hline
$\ptjet$&  $\sigma \pm {  (stat.)} \pm {  (sys.)}$ & $  \rm C_{  \rm HAD}$ \\  
$[{\rm GeV}/c]$  & [nb/(GeV/$c$)] & parton $\rightarrow$ hadron\\ \hline
54 - 62  & $(  6.67  \pm  0.15  {}^{+ 0.84 }_{ -0.75 }) \times 10^{0\ \ }$ & $ 1.132  \pm  0.104 $\\
62 - 72  & $(  2.68  \pm  0.02  {}^{+ 0.32 }_{ -0.30 }) \times 10^{0\ \ }$ & $ 1.116  \pm  0.087 $\\
72 - 83  & $(  1.04  \pm  0.01  {}^{+ 0.12 }_{ -0.12 }) \times 10^{0\ \ }$ & $ 1.100  \pm  0.072 $\\
83 - 96  & $(  3.77  \pm  0.04  {}^{+ 0.49 }_{ -0.46 }) \times 10^{-1}$ & $ 1.086  \pm  0.058 $\\
96 - 110  & $( 1.32  \pm  0.02  {}^{+ 0.19 }_{ -0.18 }) \times 10^{-1}$ & $ 1.072  \pm  0.045 $\\
110 - 127  & $(  4.18  \pm  0.04  {}^{+ 0.72 }_{ -0.65 }) \times 10^{-2}$ & $ 1.059  \pm  0.033 $\\
127 - 146  & $(  1.21  \pm  0.02  {}^{+ 0.24 }_{ -0.22 }) \times 10^{-2}$ & $ 1.047  \pm  0.022 $\\
146 - 169  & $(  2.92  \pm  0.04  {}^{+ 0.70 }_{ -0.61 }) \times 10^{-3}$ & $ 1.035  \pm  0.012 $\\
169 - 195  & $(  5.74  \pm  0.09  {}^{+ 1.65 }_{ -1.38 }) \times 10^{-4}$ & $ 1.024  \pm  0.003 $\\
195 - 224  & $(  8.49  \pm  0.31  {}^{+ 3.09 }_{ -2.42 }) \times 10^{-5}$ & $ 1.013  \pm  0.005 $\\
224 - 259  & $(  8.65  \pm  0.63  {}^{+ 4.18 }_{ -3.08 }) \times 10^{-6}$ & $ 1.003  \pm  0.012 $\\
259 - 298  & $(  5.67  \pm  1.65  {}^{+ 3.25 }_{ -2.80 }) \times 10^{-7}$ & $ 0.993  \pm  0.018 $\\ \hline\hline
\end{tabular} 
\end{center}  
\renewcommand{\baselinestretch}{1} 
\end{footnotesize}  
\label{tab:table2}   
\caption{Measured inclusive jet differential cross section  as a function of $\ptjet$ for jets in the regions 
$0.7 < |\yjet| < 1.1$, $1.1 < |\yjet| < 1.6$, and $1.6 < |\yjet| < 2.1$  (see Fig.~\ref{fig:pt}). 
An additional $5.8 \%$ uncertainty on the integrated luminosity is
not included.
The parton-to-hadron
 correction
factors, $  {\rm C}_{  \rm HAD} (\ptjet,\yjet)$, are applied to the pQCD predictions (see Fig.~\ref{fig:chad}). 
}
\end{table}



\begin{table}[htbp] 
\begin{footnotesize} 
\begin{center} 
\renewcommand{\baselinestretch}{1.2} 
\begin{tabular}{|c|c|c|c|c|c|} \hline
\multicolumn{6}{|c|}{\normalsize{ Systematic uncertainties [\%] $ (0.1 < |\yjet|<0.7) \ \  (  D=0.5)$}} \\ \hline\hline
$\ptjet$ [GeV/$c$] & jet energy scale & resolution & unfolding & $\ptjet$-spectra & $\deltapt$ \\ \hline 
54 - 62  & ${}^{+ 9.9 }_{ -9.2 }$   & ${}^{+ 2.4 }_{ -2.3 }$   & $ \pm 5.4 $ & $\pm 0.6 $ & ${}^{+ 0.8 }_{ -0.8}$\\
62 - 72  & ${}^{+ 9.8 }_{ -9.0 }$   & ${}^{+ 2.4 }_{ -2.2 }$   & $ \pm 4.8 $ & $\pm 0.6 $ & ${}^{+ 0.7 }_{ -0.7}$\\
72 - 83  & ${}^{+ 9.8 }_{ -8.9 }$   & ${}^{+ 2.3 }_{ -2.2 }$   & $ \pm 4.3 $ & $\pm 0.6 $ & ${}^{+ 0.6 }_{ -0.7}$\\
83 - 96  & ${}^{+ 9.7 }_{ -8.9 }$   & ${}^{+ 2.2 }_{ -2.1 }$   & $ \pm 3.8 $ & $\pm 0.6 $ & ${}^{+ 0.6 }_{ -0.6}$\\
96 - 110  & ${}^{+ 9.8 }_{ -9.0 }$  & ${}^{+ 2.2 }_{ -2.1 }$   & $ \pm 3.4 $ & $\pm 0.6 $ & ${}^{+ 0.5 }_{ -0.5}$\\
110 - 127  & ${}^{+ 10.0 }_{ -9.4 }$   & ${}^{+ 2.1 }_{ -2.0 }$   & $ \pm 3.1 $ & $\pm 0.6 $ & ${}^{+ 0.5 }_{ -0.5}$\\
127 - 146  & ${}^{+ 10.4 }_{ -9.9 }$   & ${}^{+ 2.1 }_{ -2.0 }$   & $ \pm 2.8 $ & $\pm 0.6 $ & ${}^{+ 0.4 }_{ -0.4}$\\
146 - 169  & ${}^{+ 11.2 }_{ -10.8 }$  & ${}^{+ 2.1 }_{ -2.0 }$   & $ \pm 2.5 $ & $\pm 0.5 $ & ${}^{+ 0.4 }_{ -0.4}$\\
169 - 195  & ${}^{+ 12.5 }_{ -11.9 }$  & ${}^{+ 2.1 }_{ -2.1 }$   & $ \pm 2.3 $ & $\pm 0.4 $ & ${}^{+ 0.4 }_{ -0.4}$\\
195 - 224  & ${}^{+ 14.3 }_{ -13.3 }$  & ${}^{+ 2.2 }_{ -2.2 }$   & $ \pm 2.1 $ & $\pm 0.3 $ & ${}^{+ 0.4 }_{ -0.3}$\\
224 - 259  & ${}^{+ 16.6 }_{ -15.0 }$  & ${}^{+ 2.4 }_{ -2.4 }$   & $ \pm 1.9 $ & $\pm 0.2 $ & ${}^{+ 0.3 }_{ -0.3}$\\
259 - 298  & ${}^{+ 19.3 }_{ -17.0 }$  & ${}^{+ 2.7 }_{ -2.7 }$   & $ \pm 1.8 $ & $\pm 0.1 $ & ${}^{+ 0.3 }_{ -0.3}$\\
298 - 344  & ${}^{+ 22.3 }_{ -19.4 }$  & ${}^{+ 3.1 }_{ -3.2 }$   & $ \pm 1.6 $ & $\pm 0.1 $ & ${}^{+ 0.3 }_{ -0.3}$\\
344 - 396  & ${}^{+ 25.7 }_{ -22.1 }$  & ${}^{+ 3.7 }_{ -3.8 }$   & $ \pm 1.5 $ & $\pm 0.2 $ & ${}^{+ 0.3 }_{ -0.3}$\\
396 - 457  & ${}^{+ 30.7 }_{ -25.5 }$  & ${}^{+ 4.5 }_{ -4.6 }$   & $ \pm 1.4 $ & $\pm 0.5 $ & ${}^{+ 0.3 }_{ -0.3}$\\
457 - 527  & ${}^{+ 39.5 }_{ -29.7 }$  & ${}^{+ 5.5 }_{ -5.6 }$   & $ \pm 1.3 $ & $\pm 1.3 $ & ${}^{+ 0.3 }_{ -0.2}$\\
527 - 700  & ${}^{+ 52.6 }_{ -37.7 }$  & ${}^{+ 7.4 }_{ -7.3 }$   & $ \pm 1.2 $ & $\pm 4.2 $ & ${}^{+ 0.3 }_{ -0.2}$\\ \hline\hline
\multicolumn{6}{|c|}{\normalsize{ Systematic uncertainties [\%] $ (0.1 < |\yjet|<0.7 ) \ \ (  D=1.0)$}} \\ \hline\hline
$\ptjet$ [GeV/$c$] & jet energy scale & resolution & unfolding & $\ptjet$-spectra & $\deltapt$ \\ \hline
54 - 62  & ${}^{+ 10.7 }_{ -9.4 }$   & ${}^{+ 2.7 }_{ -2.7 }$   & $ \pm 5.6 $ & $\pm 0.4 $ & ${}^{+ 3.5 }_{ -2.9}$\\
62 - 72  & ${}^{+ 10.4 }_{ -9.3 }$   & ${}^{+ 2.6 }_{ -2.5 }$   & $ \pm 4.9 $ & $\pm 0.4 $ & ${}^{+ 3.0 }_{ -2.6}$\\
72 - 83  & ${}^{+ 10.3 }_{ -9.2 }$   & ${}^{+ 2.4 }_{ -2.4 }$   & $ \pm 4.2 $ & $\pm 0.4 $ & ${}^{+ 2.6 }_{ -2.4}$\\
83 - 96  & ${}^{+ 10.2 }_{ -9.2 }$   & ${}^{+ 2.3 }_{ -2.3 }$   & $ \pm 3.7 $ & $\pm 0.4 $ & ${}^{+ 2.3 }_{ -2.2}$\\
96 - 110  & ${}^{+ 10.2 }_{ -9.3 }$  & ${}^{+ 2.2 }_{ -2.2 }$   & $ \pm 3.2 $ & $\pm 0.4 $ & ${}^{+ 2.1 }_{ -2.0}$\\
110 - 127  & ${}^{+ 10.4 }_{ -9.6 }$  & ${}^{+ 2.1 }_{ -2.1 }$   & $ \pm 2.8 $ & $\pm 0.4 $ & ${}^{+ 1.9 }_{ -1.8}$\\
127 - 146  & ${}^{+ 10.8 }_{ -10.1 }$ & ${}^{+ 2.0 }_{ -2.0 }$   & $ \pm 2.5 $ & $\pm 0.4 $ & ${}^{+ 1.7 }_{ -1.7}$\\
146 - 169  & ${}^{+ 11.5 }_{ -10.8 }$ & ${}^{+ 1.9 }_{ -1.9 }$   & $ \pm 2.1 $ & $\pm 0.4 $ & ${}^{+ 1.6 }_{ -1.6}$\\
169 - 195  & ${}^{+ 12.6 }_{ -11.8 }$ & ${}^{+ 1.9 }_{ -2.0 }$   & $ \pm 1.9 $ & $\pm 0.4 $ & ${}^{+ 1.5 }_{ -1.4}$\\
195 - 224  & ${}^{+ 13.9 }_{ -13.1 }$ & ${}^{+ 1.9 }_{ -2.0 }$   & $ \pm 1.6 $ & $\pm 0.3 $ & ${}^{+ 1.4 }_{ -1.3}$\\
224 - 259  & ${}^{+ 15.8 }_{ -14.7 }$ & ${}^{+ 2.1 }_{ -2.2 }$   & $ \pm 1.4 $ & $\pm 0.3 $ & ${}^{+ 1.3 }_{ -1.3}$\\
259 - 298  & ${}^{+ 18.0 }_{ -16.6 }$ & ${}^{+ 2.4 }_{ -2.5 }$   & $ \pm 1.3 $ & $\pm 0.2 $ & ${}^{+ 1.3 }_{ -1.2}$\\
298 - 344  & ${}^{+ 20.8 }_{ -18.8 }$ & ${}^{+ 2.8 }_{ -2.9 }$   & $ \pm 1.1 $ & $\pm 0.2 $ & ${}^{+ 1.2 }_{ -1.1}$\\
344 - 396  & ${}^{+ 24.5 }_{ -21.4 }$ & ${}^{+ 3.4 }_{ -3.6 }$   & $ \pm 1.0 $ & $\pm 0.2 $ & ${}^{+ 1.2 }_{ -1.1}$\\
396 - 457  & ${}^{+ 30.1 }_{ -24.7 }$ & ${}^{+ 4.3 }_{ -4.4 }$   & $ \pm 0.8 $ & $\pm 0.5 $ & ${}^{+ 1.1 }_{ -1.0}$\\
457 - 527  & ${}^{+ 38.8 }_{ -29.5 }$ & ${}^{+ 5.4 }_{ -5.4 }$   & $ \pm 0.7 $ & $\pm 1.1 $ & ${}^{+ 1.1 }_{ -1.0}$\\
527 - 700  & ${}^{+ 49.8 }_{ -37.6 }$ & ${}^{+ 7.3 }_{ -7.2 }$   & $ \pm 0.6 $ & $\pm 3.4 $ & ${}^{+ 1.0 }_{ -0.9}$\\ \hline\hline
\end{tabular}
\end{center}  
\renewcommand{\baselinestretch}{1} 
\end{footnotesize}  
\label{tab:sysktds}   
\caption{
Systematic uncertainties (in percent) on the
measured inclusive jet differential cross section  as a function of $\ptjet$, for jets in the region 
$0.1 < |\yjet| < 0.7$ and using $  D=0.5$ and $  D=1.0$ (see Fig.~\ref{fig:ktds}).  The different columns follow the discussion in Section~X.
An additional $5.8 \%$ uncertainty on the integrated luminosity is not included. 
}
\end{table}   
\clearpage


\begin{table}[htbp] 
\begin{footnotesize} 
\begin{center} 
\renewcommand{\baselinestretch}{1.2} 
\begin{tabular}{|c|r|c|} \hline
\multicolumn{3}{|c|}{\normalsize{$ \frac{ {d^2} \sigma}{ {d} \ptjet  {d} \yjet} \ (0.1 <|\yjet|<0.7) \ \ (  D=0.5)$}} \\ \hline\hline
$\ptjet$&  $\sigma \pm {  (stat.)} \pm {  (sys.)}$ & $  \rm C_{  \rm HAD}$ \\  
$[{\rm GeV}/c]$  & [nb/(GeV/$c$)] & parton $\rightarrow$ hadron\\ \hline
54 - 62  & $(  10.5  \pm  0.2  {}^{+ 1.2 }_{ -1.1 }) \times 10^{0\ \ }$  & $ 1.089  \pm  0.104 $\\
62 - 72  & $(  4.81  \pm  0.03  {}^{+ 0.54 }_{ -0.50 }) \times 10^{0\ \ }$  & $ 1.076  \pm  0.086 $\\
72 - 83  & $(  2.09  \pm  0.01  {}^{+ 0.23 }_{ -0.21 }) \times 10^{0\ \ }$  & $ 1.064  \pm  0.070 $\\
83 - 96  & $(  0.91  \pm  0.01  {}^{+ 0.10 }_{ -0.09 }) \times 10^{0\ \ }$  & $ 1.055  \pm  0.057 $\\
96 - 110  & $(   3.95  \pm  0.04  {}^{+ 0.42 }_{ -0.39 }) \times 10^{-1}$  & $ 1.047  \pm  0.047 $\\
110 - 127  & $(  1.71  \pm  0.02  {}^{+ 0.18 }_{ -0.17 }) \times 10^{-1}$  & $ 1.041  \pm  0.037 $\\
127 - 146  & $(  0.71  \pm  0.01  {}^{+ 0.08 }_{ -0.07 }) \times 10^{-1}$  & $ 1.035  \pm  0.029 $\\
146 - 169  & $(  2.76  \pm  0.02  {}^{+ 0.32 }_{ -0.31 }) \times 10^{-2}$  & $ 1.030  \pm  0.023 $\\
169 - 195  & $(  1.04  \pm  0.01  {}^{+ 0.14 }_{ -0.13 }) \times 10^{-2}$  & $ 1.026  \pm  0.017 $\\
195 - 224  & $(  3.87  \pm  0.02  {}^{+ 0.57 }_{ -0.53 }) \times 10^{-3}$  & $ 1.022  \pm  0.012 $\\
224 - 259  & $(  1.34  \pm  0.01  {}^{+ 0.23 }_{ -0.21 }) \times 10^{-3}$  & $ 1.019  \pm  0.008 $\\
259 - 298  & $(  4.26  \pm  0.04  {}^{+ 0.83 }_{ -0.74 }) \times 10^{-4}$  & $ 1.017  \pm  0.005 $\\
298 - 344  & $(  1.22  \pm  0.02  {}^{+ 0.28 }_{ -0.24 }) \times 10^{-4}$  & $ 1.015  \pm  0.002 $\\
344 - 396  & $(  3.16  \pm  0.09  {}^{+ 0.82 }_{ -0.71 }) \times 10^{-5}$  & $ 1.013  \pm  0.001 $\\
396 - 457  & $(  6.30  \pm  0.32  {}^{+ 1.96 }_{ -1.63 }) \times 10^{-6}$  & $ 1.011  \pm  0.002 $\\
457 - 527  & $(  1.01  \pm  0.12  {}^{+ 0.40 }_{ -0.31 }) \times 10^{-6}$  & $ 1.010  \pm  0.003 $\\
527 - 700  & $(  0.83  \pm  0.23  {}^{+ 0.44 }_{ -0.32 }) \times 10^{-7}$  & $ 1.008  \pm  0.005 $\\ \hline\hline
\multicolumn{3}{|c|}{\normalsize{$ \frac{ {d^2} \sigma}{ {d} \ptjet  {d} \yjet} \ (0.1 < |\yjet|<0.7) \ \ (  D=1.0)$}} \\ \hline\hline
$\ptjet$&  $\sigma \pm {  (stat.)} \pm {  (sys.)}$ & $  \rm C_{  \rm HAD}$ \\  
$[{\rm GeV}/c]$  & [nb/(GeV/$c$)] & parton $\rightarrow$ hadron\\ \hline
54 - 62  & $(  20.0  \pm  0.2   {}^{+ 2.6 }_{ -2.3 }) \times 10^{0\ \ }$  & $ 1.372  \pm  0.227$ \\
62 - 72  & $(  8.65  \pm  0.04  {}^{+ 1.1 }_{ -1.0 }) \times 10^{0\ \ }$  & $ 1.296  \pm  0.171$ \\
72 - 83  & $(  3.59  \pm  0.02  {}^{+ 0.42 }_{ -0.39 }) \times 10^{0\ \ }$  & $ 1.236  \pm  0.129$ \\
83 - 96  & $(  1.49  \pm  0.01  {}^{+ 0.17 }_{ -0.16 }) \times 10^{0\ \ }$  & $ 1.190  \pm  0.098$ \\
96 - 110  & $(  6.27  \pm   0.05  {}^{+ 0.70 }_{ -0.65 }) \times 10^{-1}$  & $ 1.155  \pm  0.075$ \\
110 - 127  & $(  2.63  \pm  0.03  {}^{+ 0.29 }_{ -0.27 }) \times 10^{-1}$  & $ 1.127  \pm  0.057$ \\
127 - 146  & $(  1.05  \pm  0.01  {}^{+ 0.12 }_{ -0.11 }) \times 10^{-1}$  & $ 1.105  \pm  0.044$ \\
146 - 169  & $(  4.04  \pm  0.03  {}^{+ 0.48 }_{ -0.45 }) \times 10^{-2}$  & $ 1.088  \pm  0.034$ \\
169 - 195  & $(  1.48  \pm  0.01  {}^{+ 0.19 }_{ -0.18 }) \times 10^{-2}$  & $ 1.075  \pm  0.026$ \\
195 - 224  & $(  5.41  \pm  0.02  {}^{+ 0.77 }_{ -0.73 }) \times 10^{-3}$  & $ 1.065  \pm  0.019$ \\
224 - 259  & $(  1.86  \pm  0.01  {}^{+ 0.30 }_{ -0.28 }) \times 10^{-3}$  & $ 1.057  \pm  0.013$ \\
259 - 298  & $(  5.77  \pm  0.04  {}^{+ 1.05 }_{ -1.00 }) \times 10^{-4}$  & $ 1.050  \pm  0.008$ \\
298 - 344  & $(  1.70  \pm  0.02  {}^{+ 0.36 }_{ -0.32 }) \times 10^{-4}$  & $ 1.045  \pm  0.003$ \\
344 - 396  & $(  4.26  \pm  0.10  {}^{+ 1.05 }_{ -0.93 }) \times 10^{-5}$  & $ 1.041  \pm  0.003$ \\
396 - 457  & $(  8.17  \pm  0.36  {}^{+ 2.49 }_{ -2.06 }) \times 10^{-6}$  & $ 1.038  \pm  0.009$ \\
457 - 527  & $(  1.39  \pm  0.14  {}^{+ 0.55 }_{ -0.42 }) \times 10^{-6}$  & $ 1.036  \pm  0.015$ \\
527 - 700  & $(  1.19  \pm  0.27  {}^{+ 0.60 }_{ -0.46 }) \times 10^{-7}$  & $ 1.033  \pm  0.027$ \\ \hline\hline
\end{tabular} 
\end{center}  
\renewcommand{\baselinestretch}{1} 
\end{footnotesize}  
\label{tab:tablektds}   
\caption{Measured inclusive jet differential cross section  as a function of $\ptjet$ for jets in the region 
$0.1 < |\yjet| < 0.7$ using $  D=0.5$ and $  D=1.0$ (see Fig.~\ref{fig:ktds}). 
An additional $5.8 \%$ uncertainty on the integrated  luminosity is
not included.
The parton-to-hadron
 correction
factors, $ {\rm  C}_{  \rm HAD} (\ptjet)$, are applied to the pQCD predictions. 
}
\end{table}   
\clearpage


\appendix
\section{Correlations of  systematic uncertainties}

The correlations among systematic uncertainties in different $\ptjet$ bins and $|\yjet|$ regions     
are studied in detail. The uncertainty on the absolute jet energy scale is 
decomposed into different sources considered independent but fully correlated across $\ptjet$ bins and $|\yjet|$ regions.
A  $\pm 1.8 \%$ 
uncertainty on the absolute energy scale, independent of $\ptjet$, results from the sum in quadrature 
of four different contributions~\cite{jetcornim}: a $\pm 0.5 \%$ uncertainty  from the calorimeter stability versus time, a $\pm 1.0 \%$ uncertainty 
due to the  modeling of the jet fragmentation, a $\pm 0.5 \%$ uncertainty  from  the simulation of the 
electromagnetic calorimeter response, and a $\pm 1.3 \%$ uncertainty from the simulation of the calorimeter response at the boundary between calorimeter towers.
Other contributions to  the absolute energy scale uncertainty come from the description of the calorimeter response to  hadrons for different 
ranges in hadron momentum~\cite{jetcornim}. 
Table~VIII shows the resulting relative contributions to the quoted systematic uncertainty on the measured cross sections 
related to the  absolute jet energy scale uncertainty.

\begin{table}[htbp] 
\begin{footnotesize} 
\begin{center} 
\renewcommand{\baselinestretch}{1.2} 
\begin{tabular}{|c|c|c|c|c|} \hline
\multicolumn{1}{|c|}{\footnotesize{$\ptjet$}} & 
\multicolumn{1}{|c|}{\footnotesize{$\ptjet$ independent}} &
\multicolumn{3}{|c|}{\footnotesize{response to hadrons}} \\  
\multicolumn{1}{|c|}{\footnotesize{[GeV/$c$]}} &
\multicolumn{1}{|c|}{\footnotesize{uncertainty}} &
\multicolumn{1}{|c|}{\footnotesize{$  p < 12$   GeV/$c$}} &
\multicolumn{1}{|c|}{\footnotesize{$12 <   p < 20$   GeV/$c$}} &
\multicolumn{1}{|c|}{\footnotesize{$  p > 20$   GeV/$c$}} \\ \hline
 54 - 62 & 90.3 & 37.8 & 15.2 & 13.5\\
 62 - 72 & 90.2 & 35.2 & 16.1 & 19.1\\
 72 - 83 & 89.9 & 31.9 & 17.0 & 24.6\\
 83 - 96 & 89.2 & 28.8 & 17.3 & 30.1\\
 96 - 110 & 88.0 & 26.0 & 16.9 & 35.8\\
110 - 127 & 86.4 & 22.7 & 16.4 & 41.9\\
127 - 146 & 84.3 & 20.0 & 15.1 & 47.7\\
146 - 169 & 82.1 & 17.2 & 14.1 & 52.6\\
169 - 195 & 79.8 & 14.6 & 12.7 & 57.0\\
195 - 224 & 77.6 & 12.5 & 11.5 & 60.7\\
224 - 259 & 75.7 & 10.7 & 10.3 & 63.6\\
259 - 298 & 73.8 &  9.1 &  9.2 & 66.2\\
298 - 344 & 72.1 &  7.8 &  8.2 & 68.3\\
344 - 396 & 70.5 &  6.8 &  7.3 & 70.2\\
396 - 457 & 69.2 &  5.8 &  6.4 & 71.7\\
457 - 527 & 68.0 &  5.0 &  5.7 & 72.9\\
527 - 700 & 66.8 &  4.2 &  5.0 & 74.2\\ \hline\hline
\end{tabular}
\end{center}  
\renewcommand{\baselinestretch}{1} 
\end{footnotesize}  
\label{tab:decompose}   
\caption{Relative contributions (in percent) to the quoted systematic uncertainty on the measured cross sections 
related to the  absolute jet energy scale uncertainty. The second column corresponds to a $\pm 1.8 \%$ 
uncertainty on the absolute energy scale, as discussed in the text. Sources are considered independent and fully correlated  in $\ptjet$ and $|\yjet|$.}
\end{table}   

\noindent
The rest of the systematic uncertainties on the measured cross sections, including that on the total integrated 
luminosity, are also assumed to be  independent and fully correlated across $\ptjet$ bins and $|\yjet|$ regions, 
except those related to the $\beta_{\rm  data}/\beta_{\rm  mc}$ ratio, for which uncertainties in different $|\yjet|$ regions are uncorrelated.

A global $\chi^2$ test is performed according to the formula 
\begin{equation}
\chi^2 = \sum_{j=1}^{76} \frac{[\sigma^{  d}_j - \sigma^{  th}_j(\bar{s})]^2}{[\delta\sigma^{  d}_j]^2 + [\delta\sigma^{  th}_j(\bar{s})]^2 } 
+ \sum_{i=1}^{17} [s_i]^2 \ , 
\end{equation}
\noindent
where $\sigma^{  d}_j$ is the measured cross section for  data point $j$,   $\sigma^{  th}_j(\bar{s})$ is 
the corresponding prediction, and  $\bar{s}$  
denotes the vector of  standard deviations, $s_i$, for the different independent sources of 
systematic uncertainty. The values for $\sigma^{  th}_j (\bar{s})$ are obtained
from the nominal NLO pQCD prediction, where $\bar{s}$ includes the uncertainty on $ \rm C_{ \rm HAD}$ but
does not consider PDF uncertainties. The uncertainty on  $ \rm C_{ \rm HAD}$ is assumed to be fully
correlated across  $\ptjet$ bins and $|\yjet|$ regions.
The sums in Eq.~(A1) run over 76 data points and 17
independent sources of systematic uncertainty, and the $\chi^2$ is minimized with respect to $\bar{s}$.
Correlations among systematic uncertainties  
are taken into account in $\sigma^{  th}_j(\bar{s})$. 
As an example, for a given source $i$, variations of $s_i$ will coherently affect all 
the $\sigma^{  th}_j(\bar{s})$ values if the corresponding systematic uncertainties are 
considered fully correlated  across  $\ptjet$ bins and $|\yjet|$ regions.



\begin{thebibliography}{99.}
\bibitem{coord} We use a cylindrical coordinate system about the beam axis in which the proton defines the positive $z$-direction,  
$\theta$ is the polar angle, and $\phi$ is the azimuthal angle. We define $E_T = E \ {  sin} \theta$,  
$p_T = p \ {  sin}  \theta$, $\eta = - {  ln(tan(\frac{\theta}{2})) }$, and   
$y =   \frac{1}{2} ln(\frac{E+p_z}{E-p_z}) $.
\bibitem{hadron}  The hadron level in the Monte Carlo generators is defined
using all final-state particles with lifetime above $10^{-11} {  s}$.
\bibitem{fn1} The electromagnetic 
and hadronic sections of a calorimeter tower are 
defined massless. For each of them, the four-momentum is constructed using its energy 
and its position with respect to the interaction point.
\bibitem{fn2} In order to reduce online trigger rates, different
prescale factors are applied to trigger paths. For a given trigger path, a prescale of 1000 would indicate that only one of every 1000 events 
satisfying the trigger requirements 
is selected.
\bibitem{fn3}A trigger tower is defined adding the information of two adjacent calorimeter 
towers.
\bibitem{met} $\ETM$ is defined as the norm of $-\sum_i E_T^i \cdot \vec{n}_i$, where $\vec{n}_i$ is the unit vector in the azimuthal 
plane that points from the beamline to the i-th calorimeter tower.
\bibitem{fn4} If considered event-by-event, $\beta$ is equivalent to 
$\ptjettest/\ptjetref$. However, Eq.~(2) is preferred since $\Delta$    
follows a Gaussian distribution while the ratio $\ptjettest/\ptjetref$ suffers from important non-Gaussian tails.
\bibitem{pthat} $\hat{p}_{t}$ is defined for two-body reactions in terms of Mandelstam variables as $\hat{p}_{t}=\sqrt{\frac{\hat{t} \cdot \hat{u}}{\hat{s}}}$.
\end{thebibliography}
\end{document}